\documentclass[a4paper,fleqn,usenatbib]{mnras}
\usepackage[T1]{fontenc}
\usepackage{ae,aecompl}
\usepackage{graphicx}
\usepackage{amssymb}
\usepackage{amsmath}
\usepackage{mathptmx}
\usepackage{epstopdf}
\usepackage{booktabs}
\usepackage{float}
\usepackage{subfig}
\usepackage{epsfig,color,ulem}
\usepackage{tabularx}

\definecolor{green}{rgb}{0.0, 0.5, 0.0}

\newcommand{\B}{{\it B}}
\newcommand{\msun}{\,{\rm M_{\odot}}}
\newcommand{\cm}{\,{\rm cm}}
\newcommand{\rad}{\,{\rm rad}}
\newcommand{\s}{\,{\rm s}}
\def\gsim{ \lower .75ex \hbox{$\sim$} \llap{\raise .27ex \hbox{$>$}} }
\def\lsim{ \lower .75ex\hbox{$\sim$} \llap{\raise .27ex \hbox{$<$}} }

\title[]{
A cocoon shock breakout as the origin of the $ \gamma $-ray emission in GW170817 }

\author[Gottlieb, Nakar, Piran \& Hotokezaka]{
	Ore Gottlieb$^{1}$\thanks{oregottlieb@mail.tau.ac.il},
	Ehud Nakar$^{1}$,
	Tsvi Piran$^{2}$,
	Kenta Hotokezaka$^{3}$
	\\
	$^{1}${The Raymond and Beverly Sackler School of Physics and
		Astronomy, Tel Aviv University, Tel Aviv 69978, Israel}\\
	$^{2}${Racah Institute of Physics, The Hebrew University of
		Jerusalem, Jerusalem 91904, Israel}\\
	$^{3}${Department of Astrophysical Sciences, Princeton University, Peyton Hall, Princeton, NJ 08544, USA}
}
\pubyear{2017}
\begin{document}
	\label{firstpage}
	\pagerange{\pageref{firstpage}--\pageref{lastpage}}
	\maketitle	
	\begin{abstract}
The short Gamma-Ray Burst, GRB170817A, that followed the binary neutron star merger gravitational waves signal, GW170817, is not a usual sGRB. It is weaker by three orders of magnitude than the weakest sGRB seen before and its spectra, showing a hard early signal followed by a softer thermal spectrum, is unique. We show, first, that the $\gamma$-rays must have emerged from at least mildly relativistic outflow, implying that a relativistic jet was launched following the merger. We then show that the observations are consistent with the predictions of a mildly relativistic shock breakout: a minute $\gamma$-ray energy as compared with the total energy and a rather smooth light curve with a hard to soft evolution. We present here a novel analytic study and detailed numerical 2D and 3D relativistic hydrodynamic and radiation simulations that support the picture in which the observed $\gamma$-rays arose from a shock breakout of a cocoon from the merger's ejecta \citep{Kasliwal17}. The cocoon can be formed by either a choked jet which does not generate a sGRB (in any direction) or by a successful jet which generates an undetected regular sGRB along the system's axis pointing away from us. Remarkably, for the choked jet model, the macronova signal produced by the ejecta (which is partially boosted to high velocities by the cocoon's shock) and the radio that is produced by the interaction of the shocked cocoon material with the surrounding matter, agree with the observed UV/optical/IR emission and with current radio observations. Finally, we discuss the possibility that the jet propagation within the ejecta may photodissociate some of of the heavy elements and may affect the composition of a fraction of ejecta and, in turn, the opacity and the early macronova light.
	
	\end{abstract}

\section{Introduction}
\label{sec:introduction}	

The advanced LIGO and advanced Virgo gravitational radiation observatories have detected the first binary neutron star merger on August 17 2017 \citep{2017GCN.21505....1S,2017GCN.21513....1S,2017GCN.21527....1S,LIGO}. The gravitational radiation event known as GW170817 revealed a merger of two low mass compact objects, whose mass range clearly puts them as neutron star rather than black hole candidates \citep{LIGO,Multi}. Remarkably,
the $\gamma$-ray satellite Fermi detected a short gamma-ray burst (GRB) about 2 seconds after the GW signal, from a location that is consistent with the localization of the GW signal \citep{2017GCN.21528....1S}. This was followed by a detection of an optical  counterpart, SSS17a \citep{2017GCN.21529....1S}, associated with the accompanying Macronova/kilonova that  was detected also in the UV  in the IR \citep{Evans17} and later on with an X-ray \citep{2017GCN.21765....1S} and Radio  \citep{Hallinan17} counterparts. The optical observations allowed the identification of a host galaxy, NGC 4993, an early-type galaxy at a distance of $\approx 40$\,Mpc.  

While tantalizing and seemingly confirming a long standing prediction of the association of sGRBs with mergers \citep{Eichler1989}, a quick look at the $\gamma$-ray observations shows that the sGRB 170817A which was discovered in association with GW170817  is not a regular sGRB. 
While its fluence, $(2.8\pm 0.2) \times10^{-7}$ erg cm$^{-2}$,  and duration, $ 2 \pm 0.5$ s,  are similar to other observed sGRBs, when taking into account its distance of $\sim 40$ Mpc, we find that its total isotropic equivalent energy, $E_{iso}=(5.35 \pm 1.26 ) \times 10^{46}$ erg  is smaller by three orders of magnitude than the weakest sGRB whose energy was measured so far and by four orders of magnitude than typical sGRBs \citep[see e.g.][for a review]{nakar2007}. 
Furthermore, the pulse is made of two parts with a very different spectrum.  The spectrum of the main
pulse (T$_0$-0.320 s to T$_0$+0.256s)  is best fit with a Comptonized function, with a power-law index
of $ -0.62 \pm 0.40$ with peak energy $E_{peak}$ = 185 $ \pm $ 62 keV. 
The main pulse is followed by a weaker tail  extending from T$_0$+0.832 s to T$_0$+1.984 s, whose  spectrum is softer. It is  well-fit by a black body with $kT = 10.3  \pm  1.5$ keV \citep{2017GCN.21528....1S,Multi}. The fluence ratio between the two components is roughly 2:1.  This two component structure is certainly not common in sGRBs and as far as we know it has not been observed before. In fact, it seems what was discovered here is a new type of gamma-ray bursts -  a {\it low luminosity short} GRB, denoted hereafter {\it lls}GRB. 

Compactness arguments \citep{Kasliwal17}  reveal that the observed $\gamma$-rays must have been produced in a mildly or fully relativistic outflow with $\Gamma \gtrsim 2.5$. At the same time \cite{Kasliwal17}  showed  that three simple scenarios that come to ones mind: a very weak relativistic jet pointing towards us;  an off-axis emission of a strong GRB pointing along the rotation axis ($30^{\circ}$ away from us); a structured jet with a broad weak wing, are highly unlikely. We must therefore turn to something else. As we show in \S \ref{sec:jet} the fact that the emitting region must be moving relativistically cannot arise in a spherical explosion and therefore the merger must involve a relativistic jet. Numerical simulations have shown long ago \citep[e.g.][]{Davies1994,Ruffert1996,Rosswog1999,Bauswein2013,Hotokezaka2013,Sekiguchi2015, Radice2016} that neutron star mergers are accompanied by a significant amount of mass ejection. The observations of the macronova accompanying GW170817 confirms this idea putting a lower limit of $0.02 M_\odot$ on the mass ejected in this event  \citep{Kasliwal17}.  
Any relativistic jet will have to path through this mass (see \citealt{Nagakura2014,Murguia2014,Duffell2015,Lazzati2017a,g17} for jet propagation studies).  Regardless of the question whether the jet penetrates the ejecta or is choked within it,  it will produce an energetic cocoon. {We have shown in the past that the cocoon can generate a mildly relativistic outflow at a wide angle \citep{g17}. Here we suggest that the observed $\gamma$-rays are produced when  the shock driven by this mildly relativistic cocoon breaks out of the ejecta.  This process is drastically different from the one that takes place in regular sGRBs. In fact, we suggest that there are similarities between the physical mechanisms (but not in the astronomical scenario) deriving this  {\it lls}GRB and 
those that take place in regular (long) low-luminosity GRBs ({\it ll}GRBs.)}

A shock breakout of the cocoon provides a natural explanation to two puzzling properties of the observed pulse. First, the energy emitted in $\gamma$-rays in our direction is only a minute fraction, $\sim 10^{-4}$ (isotropic equivalent) of the total kinetic energy seen in the outflow ($ \gsim 10^{51}$ erg). Although a jet that can propagate a significant way through the ejecta, and thus any cocoon that it produces, is expected to carry much more energy than the one we observe in $\gamma$-rays. This is unlike regular GRBs, which are very efficient in $\gamma$-ray production. However, it is a natural property of shock breakout emission, where the radiation is coming from a very narrow layer where the optical depth upon breakout is low enough, while the bulk of the energy is in highly optically thick material and is therefore hidden (this energy is released only much later after it suffers significant adiabatic loses due to expansion). Similar bulk to breakout energy ratios are seen in breakout candidates such as SN2008D \citep{soderberg2008} and low-luminosity long GRBs \citep[e.g.,][]{Kulkarni1998,Soderberg2006}. Second, one of the predictions of a relativistic shock breakout from a star is an initial hard pulse in $\gamma$-rays,  followed by a softer (typically X-ray) tail that carries a comparable amount of energy \citep{nakar2012}. Finally, the model we present here reproduces all the main features of the observed $\gamma$-rays, the total energy, the duration, the spectrum (and its variation with time) and the 2\,s time delay between the gravitational waves and the $\gamma$-rays.

We present here a new order-of-magnitude analytic model for a shock breakout from an expanding ejecta and highlight the differences from a shock breakout from a stellar envelope. We then carry out the first detailed numerical calculations of a cocoon shock breakout in sGRBs, including the time evolution of both the luminosity and the spectrum, as well as the dependence on the viewing angle. 
The structure of the paper is as follows. 
We begin in \S \ref{sec:jet}  demonstrating using compactness  that, although we did not observe a regular sGRB,  the emitting region  must have had a relativistic jet. 
{We discuss our model in \S \ref{sec:model}. Our numerical setup, which deals with the hydrodynamic of an unmagnetized jet, and the calculation method of the breakout signal are presented in \S \ref{sec:nsimulation} . A companion paper \citep{Bromberg2017} discusses relativistic MHD simulations for the propagation of a magnetic dominated jet}. We present the configuration of a choked jet and the resulting $\gamma$-rays, UV/optical/IR and radio signals  in \S \ref{sec:choked}. A  shock breakout from a cocoon produced by a successful jet, that emerges and powers a regular sGRB pointing away from us, can also produce the observed $\gamma$-rays. We describe such a system in \S \ref{sec:succseful}.  
{Before concluding we briefly discuss in \S \ref{sec:photodissociation} the possibility of dissociation of nuclei within the shocked ejecta matter via photodissociation or due to interaction with relativistic neutrons. We show that in the setups we have considered these effects can most likely be ignored, but that it may be important in other scenarios or other events.  Detailed calculations of this process will be discussed elsewhere.} 
We conclude with implications and observational predictions in \S \ref{sec:conclusions}.

\section{Compactness and Relativistic Motion } 
\label{sec:jet}

{Remarkably, in spite of the weakness of this burst  its 
{$\gamma$-ray emission} clearly demonstrates that the first non-thermal component of this burst  must have involved relativistic motion towards us. The simple compactness arguments show that if the source would have been Newtonian, its optical depth due to pair production would have been too large to be consistent with the observations. The optical depth of the emitting region can be estimated as \citep{nakar2007,Kasliwal17}: 
\begin{eqnarray} 
&&\tau \approx \frac {3 \sigma_T f(\Gamma) E_{iso,1} }{\bar E_\gamma 4 \pi (c T_{90,1} \Gamma^2 )^2 } \approx
 \nonumber \\ 
&& 6 \times 10^7  f(\Gamma) \Gamma^{-4} \frac {E_{iso,1}}{3 \times 10^{46} ~{\rm erg}} \Big(\frac { \bar E_\gamma}{150 ~{\rm keV}}\Big)^{-1} \Big(\frac{T_{90,1} }{0.6 ~{\rm s}}\Big)^{-2}  \ , 
\label{eq:comp}
\end{eqnarray}
where  $\Gamma$ is the Lorentz factor of the emitting region  $E_{iso,1}$  is the isotropic equivalent energy of the first component,  $\bar E_\gamma$ 
is the average energy of its photons,  $T_{90,1}$ its duration  and $f(\Gamma)$ is the fraction of photons of this component that are above the $e^+ e^-$ production threshold. In GRB 170817A this is a dominant factor as the spectrum drops exponentially  $f(\Gamma) \sim \Gamma ^{-0.6} \exp [-\Gamma m_e c^2 /125 {\rm keV}]$. Now, if the source is Newtonian then $\tau \sim 10^6$,
implying first a diffusion time that is too long unless the emission comes from an unrealistically thin shell ($<10^5$ cm while the emitting region is at radius $\sim 10^{10}$\,cm) and even if such a shell existed the extremely high density of pairs in it  ($\sim 10^{26}\, {\rm cm^{-3}}$) would have  generated enough photons during  the diffusion time to obtain thermal equilibrium at a temperature much lower than the observed $\bar E_\gamma$.}

{Thus the source cannot be Newtonian. However the exponential cut-off in the spectrum makes $f({\Gamma})$ extremely sensitive to the Lorentz factor, and therefore even a mildly relativistic Lorentz factor of 2.5 is enough to make the source optically thin. We therefore conclude that the emitting region must have been moving with at least a mildly relativistic velocity. }
An immediate implication of this result is that a relativistic jet must have been involved.  It is clear from the optical observations that the bulk of the outflow is moving at $0.1$-$0.2c$ \citep{Kasliwal17}. This is much slower than the required speed. The spherical ejecta can have a fast tail, and in fact we argue later that there is one (with   $\sim10^{-7} M_\odot$ moving at a typical velocity of $\sim 0.6-0.8c$). {However, the highest velocity component of the spherical ejecta material is ejected over a very short time scale {$\sim 10$ ms} and at a radius $\lesssim 10^7$ cm, while the duration and the delay of the $\gamma$-ray signal implies that it should have been radiated at a radius of $\sim 10^{11}$ cm . At this radius the ejecta has already lost all its internal energy due to adiabatic expansion and it moves homologously.} Therefore, while this material contains enough energy to power the observed $\gamma$-rays, there is no internal mechanism to this outflow which can dissipate the bulk kinetic energy to internal energy and produce $\gamma$-rays at this point. The most plausible way to dissipate this energy is via energy emitted from the post merger compact object. { There is $\ge 0.02 M_\odot$ between the central engine and the ejecta front, which are moving at $\lesssim0.2$c also after the energy that generates the $\gamma$-rays is deposited in the outer layers.  Therefore, this energy cannot be channeled out by a spherical outflow and a relativistic jet seems to be necessary.}

\section{A Relativistic shock breakout from an expanding ejecta} 
\label{sec:model}

As mentioned earlier, the optical/IR observations \citep{Kasliwal17} clearly indicate the existence of a few percent of a solar mass that have been ejected during this merger. Most of this material is moving at $0.1$-$0.2c$. A regular sGRB jet would have had to penetrate through this ejecta before emerging and producing the $\gamma$-rays. A second possibility is that the central engine did produce a jet but this jet was choked as it could not penetrate through the ejecta {(either due to a low power{, short duration} or a large opening angle).} In fact, just a short while ago \cite{Moharana2017}  inspected the duration distribution of sGRBs and 
suggested an evidence for a significant number of sGRBs with choked jets. 
\cite{g17} have shown that a jet that successfully penetrates the ejecta can drive a mildly relativistic cocoon which expands over a large angle. Here we show that a choked jet is capable of doing that as well, if it is powerful enough. Thus, both scenarios generate a mildly relativistic shock breakout from the expanding ejecta. Below we discuss the properties of the signal that such a breakout produces.

The jet drives a radiation mediated shock into the expanding ejecta. 
 The jet drives a radiation mediated shock into the expanding ejecta. Since the jet is relativistic and it is expaning into a rather dilute plasma its head velocity is at least mildly-relativistic and hence the shock is also mildly relativistic, regardless of the details of the density profile.

The shock propagates as long as the optical depth to infinity, $\tau$, is large enough to sustain its width. Once $\tau$  drops below this point, the radiation in the shock transition layer escapes to infinity and the shock `breaks out'. Following the breakout the radiation that was deposited by the shock starts diffusing out of the expanding gas, generating the so called `cooling emission'.  The theory of shock breakout and the emission that follows was studied mostly in the context of a shock that propagates in a stellar envelope, both at Newtonian velocities \citep[e.g.][]{Colgate1974,chevalier1976,Falk1978,Matzner1999,nakar2010} and relativistic velocities \citep{nakar2012}. 

Breakout from an expanding ejecta with an edge, as expected in this case, was not considered so far in the literature. It is different than a breakout from a static stellar envelope in two important ways. First, there are two characteristic velocities instead of one, the shock velocity  as seen in the lab frame, which determines the boost of the emission to the observer, and the shock velocity in the upstream (i.e., ejecta) frame, which determines the shock strength and can be significantly different from the former. Second, unlike the breakout from a stellar edge, the diffusion length scale (i.e., the scale over which the optical depth changes significantly) and the hydrodynamic length scale (i.e., the scale over which the density, pressure and velocity change significantly) can be different. As explained below these differences can have an important effect on the observed emission. 

The resulting light curve and spectrum can depend on details such as the ejecta profile and the density near its edge. We defer a detailed  study of the signal to a future study. {Here we discuss first the dynamical evolution which is common to all breakout scenarios followed by a discussion of the spectrum that this evolution dictates. We then discuss the main robust features that the general dynamical evolution induce for a wide range of the breakout parameters. We also derive an order of magnitude estimate of the breakout parameters that are needed in order to produce the observed signal, showing that such parameters exist, that they are over constrained (i.e., more observables than free parameters) and very reasonable.}

\subsection{Planar and spherical phases}
We consider a breakout of a spherical relativistic shock, with a Lorentz factor $\Gamma_s$ and velocity $v_s=\beta_s c$ as seen in the lab frame, from ejecta that expands with a maximal velocity $\beta_{ej,max}$, that can be Newtonian or mildly relativistic. The shock velocity as seen in the ejecta frame, $\beta_{s}'=(\beta_s-\beta_{ej,max})/(1-\beta_s\beta_{ej,max})$, can be mildly relativistic or even sub-relativistic. The shock breakout takes place at a radius $R_{bo}$ from a layer with an optical depth $\sim c/v_{sh}'$, which we denote as {\it the breakout layer}. In general this layer is much narrower in the lab frame than the causality scale $R_{bo}/\Gamma^2$.  Therefore, the hydrodynamics, and as a result the observed signal, has two phases: planar and spherical. The planar phase starts right after the breakout and it lasts a duration $\sim R_{bo}/c$  in the lab frame ($\sim R_{bo}/c\Gamma_s^2$  in the observer frame), namely until the breakout layer doubles its radius. During this phase the radiation in the breakout layer, as well as in layers with higher optical depth which may be important (as we show later), evolve on time scales much shorter than $R_{bo}/c$. During this phase the radius can be approximated as a constant. In the spherical phase everything evolves on a single time scale, $R_{bo}/c$ in the lab frame (however, see also \citealt{Yalinewich2017} for the case of an acceleration in the spherical phase).  Below we discuss shortly each of this phases.

{\it  Planar phase: }The difference between a breakout from a star and a breakout from a relativistic ejecta is most important in the {  planar phase}. In a star, during this phase,  a fluid element has  only a single length scale, the initial distance from the stellar edge. Consequently,  the hydrodynamical scale and the diffusion scale are similar and the time it takes the breakout shell to expand is similar to the time it takes the radiation to diffuse through the breakout layer. This implies that during the entire planar phase only the breakout shell releases radiation to the observer (see \citealt{nakar2010} for a detailed discussion). In an expanding ejecta,  the width of the breakout layer can be much smaller than the hydrodynamical scale which we expect to be of order $R_{bo}/\Gamma_s$ in the fluid rest frame. In this case the breakout layer does not expand significantly over a diffusion time scale and photons from  inner layers have time to diffuse out during the planar phase. This has two effects: First, it increases the luminosity of the planar phase and second it modifies the spectrum since, as discussed below, the temperature of photons drops with the time it takes them to diffuse, and thus the spectrum of the planar phase is non-thermal, composed of the sum of radiation of different temperatures. Note that due to light-travel-time effects the emission from the planar phases of all the contributing layers is smeared over the angular time and seen together over a duration of $\sim R_{bo}/2c\Gamma_s^2$.

During  the {\it spherical phase}  the radius increases, the optical depth  decreases  and emission from inner layers diffuses out more efficiently.  Photons from the spherical phase are emitted after the expanding gas doubled its radius, namely they are observed at a time $\sim R_{bo}/2c\Gamma_s^2$ after the breakout emission. Therefore, they  are  not mixed with the planar phase photons. The layers that radiate during the spherical phase have much more time to cool and get closer to thermal equilibrium (see below). They also go through some adiabatic expansion and cooling. In addition, the angular time and the dynamical time are now comparable, implying that angular smearing does not mix much radiation with different temperatures. Therefore the spectrum at the beginning of the spherical phase is expected to be much softer than in the planar phase, and it is much more well defined by a single temperature. The spherical phase, that is often referred to as the cooling emission phase,  continues for a long time after the breakout. During this phase both the luminosity and the temperature drop with time. The luminosity drops gradually, while the temperature may drop quickly at first as the emitting layers are out of thermal equilibrium at the beginning of the spherical phase (see \citealt{nakar2010} for details).

{The planar and spherical phases are well separated in the observed signal when the breakout is spherical, as assumed above. If the breakout is oblique and takes place at different radii for different angles, then angular smearing can mix the emission of planar phase at one angle with the emission of spherical phase from another. If this smearing is strong then the spectrum still shows a hard to soft evolution, possibly with two well separated components, but it will remain non-thermal during the entire evolution.}

\subsection{The observed spectrum}
The observed spectrum depends on the radiation temperature in the diffusing gas. When the shock is fast enough ($\gtrsim 0.05c$ in the upstream frame) the radiation behind the shock falls out of thermal equilibrium \citep{weaver1976,katz2010,nakar2010}. 
The temperature\footnote{We  use the term  temperature also when  the gas  is out of  thermal equilibrium to  denote the typical photon  energy} is determined then by the ability of the  gas to generate photons during the  available time, which becomes  less efficient when the  shock is faster.  As a result the  radiation temperature behind the  shock rises sharply with the  shock velocity, and at $v_s' \approx 0.5c$ it reaches $\sim 50$ keV at which pair production becomes important. Once pairs are  produced efficiently the photon production rate rises as  well and the exponential dependence of the  number of pairs   on the  temperature serves as  a thermostat  that sets the temperature behind  the shock  at $\sim
100-200$\,keV for a relativistic shock, regardless of its Lorentz factor
\citep{katz2010,budnik2010,nakar2012,Granot17}. 

The   temperature evolution after the breakout depends  on the
pair loading  of the shock.  If the shock is  not too fast  ($v_s' \lesssim 0.5c$)  and pairs are negligible  then the
radiation is released upon the  breakout and the observed temperature is the gas temperature  behind the shock times its
Lorentz factor as seen  in the observer frame, $\sim \Gamma_s$.  If the shock is loaded by  pairs ($v_s' \gtrsim 0.5c$)
then the radiation  remains trapped until its rest frame  temperature drops to $\sim 50$ keV. At this stage the pairs annihilate and
the photons  are released.  In that  case the observed  temperature is  $\sim 50$ keV  times the  Lorentz factor  of the
radiating gas \citep{nakar2012}.

The radiation that diffuses to the observer following the shock breakout spends more time trapped in the gas and therefore the gas has more time to generate photons to share the internal energy, thereby reducing the radiation temperature. Quantitatively the temperature of the diffusing radiation can be estimated using the approximation presented in \cite{nakar2010}, that followed the derivation of \cite{weaver1976}. This approximation solves for the gas temperature, $T$, at which the number density of generated photons that can be coupled to the gas energy, $n_{ph}$, is enough to share the entire gas energy density, $\epsilon$, namely $\epsilon = n_{ph} 3 k_b T$, where $k_b$ is the Boltzmann constant. For that we first estimate the production rate of photons that can share the gas energy, namely photons that are emitted at the gas temperature $T$ or that can be Comptonized to that temperature in the available time. Since the typical temperatures are $\gtrsim 10$\,keV we assume that the gas is fully ionized. We consider photon production by free-free and bound free. We calculate the free-free photon production below and approximate the bound-free photon production rate to be comparable.  The free-free production rate of photons with an energy $\sim k_b T$ by a gas with a mixed composition of heavy nuclei is then \cite[e.g.,][]{nakar2010,sapir2013}
\begin{equation}
	\dot{n}_{ph,ff} \sim 3.5 \times 10^{36} \rho^2 T^{-0.5} \frac{\left< z^2 \right>\left< z \right>}{\left< A \right>^2} {\rm~cm^{-3}s^{-1}},
\end{equation} 
where $\rho$ and $T$ are in c.g.s, $z$ and $A$ are the atomic and mass numbers. The brackets mark an average over the gas and therefore depend on its composition. Here we approximate $\left(\left< z^2 \right>\left< z \right>/\left< A \right>^2\right)=10$ for r-process material. The minimal frequency,$\nu_{min}$, from which photons that are emitted by free-free at $h\nu \ll k_b T$ can be Comptonized to $T$ on time, satisfies \citep{weaver1976,nakar2010,sapir2013}
\begin{equation}
	y_{max} = \frac{k_b T}{h \nu_{min}} \approx 500 \left(\frac{\rho}{10^{-9}{\rm~ gr~cm^{-3}}}\right)^{-1/2} \left(\frac{T}{1 {\rm~keV}}\right)^{9/4} \left(\frac{\left< A \right>}{\left< z^2 \right>}\right)^{1/2}, 
\end{equation}
where we approximate $\left(\left< A \right>/\left< z^2 \right>\right)^{1/2}=5$ for r-process material.
This increases the number of photons that can share the gas energy by a factor \citep{nakar2010}
\begin{equation}
	\xi \approx \frac{1}{2} {\rm ln}[y_{max}](1.6+{\rm ln}[y_{max}])  \ .
\end{equation}
The temperature in the diffusing gas is then found by solving the implicit equation 
\begin{equation}\label{eq:Trad}
	\epsilon=6 k_b T \dot{n}_{ph,ff} \xi t,
\end{equation}
where $t$ is the time passed in the shocked gas rest frame since the crossing of the shock and the release of the photons to the observer. A factor of 2 on the r.h.s accounts for the bound-free photon production which we approximate to be comparable to free-free. This estimate ignores pairs, which is appropriate for $T\lesssim 50$ keV. If $T > 50$ keV then pairs prevent the diffusion of photons until the temperature drops to $\sim 50$ keV. 

A comparison of the results obtained from equation \ref{eq:Trad} with the dynamical simulation of \cite{sapir2013} shows that the temperature during the breakout from a stellar surface is accurate to within a factor of $\sim 2-3$.

\subsection{General properties}
The two phases described above define several general properties  that  are common to the breakout signal over a wide range
of configurations, including  both a breakout from an expanding  ejecta and from a  stellar envelope. First, the breakout
layer and all other layers, if there are any, that radiate during the planar phase contain only a very small fraction of the  total internal energy  
of the expanding gas.  As a result the  breakout pulse
contains a  very small fraction  of the total  explosion energy. Some  examples are the  emission from {\it ll}GRBs  in which  the
$\gamma$-ray energy  is only $10^{-3}  - 10^{-4}$ of  the total  explosion energy \citep[e.g.][]{Kulkarni1998,Soderberg2006}  and SN2008D in which   the energy
carried by  hard X-rays is  $\sim 10^{-5}$ of  the total SN  kinetic energy  \citep{soderberg2008}. Second, the  light curve
cannot be highly variable. It  may have some structure, especially upon the transition from  the planar to the spherical
phase, but it  cannot produce the high variability observed  in most short and long GRBs  \citep[e.g.,][]{nakar2002a,nakar2002b}. The shape of
the light curve upon  the transition from the planar to  the spherical phase depends on how  different are the diffusion
and the dynamical scales and on whether the breakout is fully  spherical or if it takes place at slightly different radii at
different angles. If the scales  are similar and the breakout is spherical (such as in  a spherical explosion in a star),
the planar phase  produces a well defined  bright pulse that is followed  by a fainter more  slowly evolving spherical phase
emission. Otherwise, the transition between these phases becomes smoother. Third, regardless of how smooth is the  transition of the luminosity from the planar to
the spherical phase, the spectrum changes significantly. The  peak of the emission is characterized by a relatively hard
spectrum that is a composition  of emission at different temperatures, while upon the  transition to the spherical phase
the spectral peak energy drops significantly and the temperature becomes more well defined. {For a more spherical  breakout the spectral transition is sharper and the temperature after the transition is more well defined.}
An example of these properties can
be seen in the detailed solution of a relativistic breakout from a static stellar envelope presented in \cite{nakar2012} {and in the results of the numerical simulations that we present here.}

\subsection{Order of magnitude estimates}
We are interested in finding an order of magnitude estimate of the properties of a shock breakout that can produce a signal that is consistent with the observed $\gamma$-rays of GRB 170817A. Namely, what are $\Gamma_s$, $\beta_{ej,max}$, $R_{bo}$ and the mass from which photons diffuse during the planar phase, $m_{bo}$, that generate a $t_{bo,obs} \sim 0.5$\,s long $\gamma$-ray pulse during the planar phase with an energy of $E_{bo} \sim 3 \times 10^{46}$ erg and a typical photon energy of  $\sim 100-150$ keV  at a delay of $\sim 2$\,s with respect to the merger time. {Upon the transition to the spherical phase the temperature should drop to $\sim 10$ keV. Note that altogether there are four breakout parameters and five observables, so the problem is over constrained and there is no guarantee that there is a viable solution.}

 The observed temperature implies that the breakout velocity in the observer frame cannot be Newtonian.  Consequently  the velocity in the ejecta rest frame   is  also at least $v'_s \gtrsim 0.5c$. Thus, the observed temperature satisfies $T \sim 50 \Gamma_{bo}$ keV implying that {the shock  must be mildly relativistic with}
\begin{equation}
	\Gamma_{bo} \approx 2-3 \ . 
\end{equation}
Note that if the shock had been  mildly relativistic also in the upstream frame (i.e., $v'_s \gtrsim 0.7c$ ), the gas would have accelerated significantly after the crossing of the shock and before the photons are released, so $\Gamma_{bo}$ would have been significantly larger than $\Gamma_s$ (see \citealt{nakar2012} for details). We  assume that this is not the case and $\Gamma_{s} \approx \Gamma_{bo}$ and verify consistency later. This breakout Lorentz factor together with the duration determines the breakout radius 
\begin{equation}
	R_{bo} \approx 2 c t_{bo,obs} \Gamma_{bo}^2 \sim 2 \times 10^{11} {\rm~cm} \ .  
\end{equation}
This radius is about 10 light seconds and it takes the shock about 11 s to travel from the source to the breakout radius. Thus, in order to obtain the observed delay the jet should be launched about a second after the merger.  The outermost part of the ejecta  has 2 seconds (denoted by $\delta t$) longer to expand than $R_{bo}/c$. Thus, it  must have a velocity 
\begin{equation}
	\beta_{ej,max} \approx \frac{R_{bo}}{R_{bo}+ c \delta t } \approx 0.7 \ . 
\end{equation}
If the shock Lorentz factor is $\Gamma_s \approx 2-3$ then this ejecta velocity { satisfies, consistently } $v'_s \approx 0.5c$. Pair production is marginal and the gas does not accelerate significantly after shock crossing, namely the assumption $\Gamma_s \approx \Gamma_{bo}$ is satisfied.

The parameters that we find above determine the energy released during the breakout with limited freedom which depends on the mass carried by the fast tail of the ejecta (i.e., at ejecta velocity $\gtrsim 0.5-0.6c$), $m_{tail}$. 
The minimal tail mass that is needed for a breakout to take place at $R_{bo}$, with $\beta'_s \approx 0.5$ is 
\begin{equation}
	m_{bo,min} \approx \frac{4 \pi R_{bo}^2}{\kappa \beta'_s} \approx 4 \times 10^{-9} \msun \ , 
\end{equation} 
where we use $\kappa=0.15\, {\rm cm^2/gr}$ as expected for fully ionized heavy elements. If $m_{tail}=m_{bo,min}$ then the width of the breakout layer in its rest frame, after it is shocked, is $\sim R_{bo}/\Gamma_s$ and the dynamical time for its expansion is comparable to the photon diffusion time through it. Therefore only this layer radiates during the planar phase. If $m_{tail}>m_{bo,min}$ then the mass which contributes to the emission during the planar phase, and thus to the energy in the initial $\gamma$-ray pulse, grows as\footnote{ We find that by equating the diffusion time through the mass $m_{bo}$ with the dynamical time in the gas frame $R/c\Gamma_s$, assuming that $m_{bo}$ is spread over a width $R m_{bo}/\Gamma_s m_{tail}$  }  $m_{bo}=(m_{bo,min} m_{tail})^{1/2}$. With a shock velocity of $\beta'_s \approx 0.5$ the internal energy in the shocked breakout mass is $\approx 0.2 m_{bo} c^2$, implying that the energy of the initial breakout pulse is:
\begin{equation}
	E_{bo} \approx  0.2 m_{bo} c^2 \Gamma_s \approx 4 \times 10^{45} \left(\frac{m_{tail}}{4 \times 10^{-9} \msun}\right)^{1/2} {\rm erg}.
\end{equation}
The observed emission implies $m_{tail} \sim 4 \times 10^{-7} \msun$. 

Here we find that a fast ejecta tail is needed for a shock breakout to explain the observed $\gamma$-rays. However, the existence of such tail was suggested in the past based on theoretical models of the merger. A fast tail of the dynamical ejecta with $v_{ej}\gtrsim 0.6c$ is likely to arise
from the interface of the merging neutron stars when the shock at the interface breaks out from the surface of the merging object \citep{Kyutoku2014}. Although it is hard to resolve numerically a small amount of fast moving components, some numerical simulations  suggest that such a fast tail exists \citep{Bauswein2013, Hotokezaka2013} \footnote{   While this paper was refereed \cite{Hotokezaka2018} analysed the highest resolution numerical simulations available of neutron star mergers \citep{Kiuchi2017}, finding a fast tail component that is similar to the one we use here.}
and that it can contain as much as $\sim 10^{-5}M_{\odot}$. Note that this amount depends
on the fate of the central object after the merger and on the neutron star equation of state, e.g., more compact neutron stars eject more fast components and the amount of the shocked ejecta is significantly reduced when
the merging neutron stars immediately collapse to a black hole \citep{Hotokezaka2013}.

Finally, we can estimate using these parameters what will be the characteristic temperature at the beginning of the spherical phase, namely, during the softer emission that follows the initial pulse. The rest frame density behind the shocked gas is roughly $\rho_s \sim  m_{tail} \Gamma_s/(4 \pi R_{bo}^3) \approx 2 \times 10^{-8} {\rm gr~cm^{-3}}$. The pressure, assuming $\beta'_s \approx 0.5$ is $p_s \approx 0.05 \rho_s c^2 \approx 10^{12} {\rm~erg~cm^{-3}}$ and the diffusion time, as measured in the gas rest frame, at the beginning of the spherical phase is $\approx R_{bo}/c \Gamma_s \approx 2.5$ s. Plugging these values to equation \ref{eq:Trad}, we obtain a rest frame temperature of $\sim 2.5$ keV and an observer frame temperature of about $7$ keV.

\section{Numerical simulations}
\label{sec:nsimulation}

The order of magnitude estimates, described earlier, demonstrate the potential of the model. We turn now to numerical simulations in order to provide a quantitative example which can be compared with the observations. {We begin with  relativistic hydrodynamic simulations.}
Then we post process the  hydrodynamic results  to obtain the observed $\gamma$-rays. 

\subsection{The relativistic hydrodynamic simulations}
We have carried out relativistic hydrodynamic numerical simulations of jet propagation and cocoon formation within the ejecta. We focused on a choked jet since it has the potential to explain also the blue optical signal seen during the first day \citep{Kasliwal17}. These simulations have used 2D axisymmetric geometry. 
We also carried out a single simulation of a successful jet to verify that it can also produce a significant $\gamma$-ray signal from the breakout of its cocoon. The early phase of this simulation required 3D. 

We used the public code PLUTO \citep{Mignone2007}, with an HLL Riemann solver and a third order Runge Kutta time stepping. Throughout the simulations we apply an equation of state with a constant adiabatic index of $ 4/3 $, as appropriate for a radiation dominated gas. We neglect gravity, as the gravitational dynamical times are longer than the typical interaction timescales.

\subsection{The $\gamma$-ray emission}
We use the results of the hydrodynamical simulations to calculate the shock breakout emission, assuming that diffusion of photons is radial, namely using the quasi-spherical symmetry approximation for photons diffusion. Following the shock breakout we find at each time step  and each angle with respect to the jet axis, $\theta$, the location within the flow from which photons can diffuse to the observer, which we term  {\it the diffusion depth}.   
The photons escape to infinity (i.e., after crossing the diffusion depth)  from the location where $\tau=1$.
For each time step and from each angle we emit the photons {from the region that crossed the diffusion depth in the last time step (along the same angle) to the observer. The observed emission is found with a proper account for the light travel time and the Lorentz boost, according to the conditions at $\tau=1$.} To find the diffusion depth as a function of time  we first identify the lab frame time of the shock breakout at this angle, $t_{bo}(\theta)$. Upon the breakout the internal energy that is in the breakout layer ($\tau=c/v'_s$) is released to the observer. After the breakout we find the diffusion depth by equating the time since the breakout to the diffusion time {(see Appendix for details)}. To calculate the spectrum which is radiated from a given diffusion depth we use equation \ref{eq:Trad} and find $T$, where the time for photon production is the time since the shock crossed the mass element at the diffusion depth and the hydrodynamical parameters are those that are at the diffusion depth at the time the photons are released. If the temperature obtained by this equation is higher than $50$\,keV we set $T=50$\,keV. The emitted spectrum is assumed to be a Wien spectrum, as expected before the radiation archive thermal equilibrium. The observed spectrum at any observer time, however, is obtained by integrating over the emission from the entire equal arrival time surface and is therefore not necessarily a Wein spectrum.

\section{Choked jet}
\label{sec:choked}

\cite{Kasliwal17} carried out numerical simulations which have shown that the interaction of a choked jet with the ejecta can simultaneously contribute to optical emission seen during the first day and  generate a shock breakout that produces a $\gamma$-ray signal. \cite{Kasliwal17} did not calculate the $\gamma$-ray emission directly form the simulations, instead they have shown that the breakout parameters are consistent with those that the order of magnitude estimates predict to produce a $\gamma$-ray signal with roughly the same energy, duration, hardness and delay as the one observed in GRB 170817A. Here we add a calculation of the $\gamma$-rays directly from the simulation (following the method explained above) in order to verify the order of magnitude estimates, including the predicted hard to soft evolution. The goal of this study is not to carry out a thorough scanning of the entire phase space in a search for an exact fit to the observed signal. Instead, we scan a relatively narrow part of the phase space to find if light curves with general properties that are similar to those observed can be generated. 
 
The observed signal depends on the mass and velocity distributions of the ejecta as well as on the jet luminosity, opening angle and duration. It also depends on the delay between the merger time, at which the ejecta starts expanding, and the jet launch. {The optical emission (and theoretical predictions) shows that the bulk of the mass is moving at $0.1$-$0.2c$. The order of magnitude estimates require that a very small fraction of the fast ejecta tail mass would extend up to $v_{ej,max} \approx 0.6-0.8c$, and similar amounts of mass are found to move at these velocities by some theoretical models  \citep{Bauswein2013,Hotokezaka2013,Kyutoku2014,Hotokezaka2018}.  Therefore,} we consider an ejecta density profile that is composed of two components, a slow core in which $ M_c $ containing a few percent of $M_\odot$, and a fast low-mass tail in which $ M_t $ containing a few percent of $ M_c $. The optical emission during the first day depends on the delay of the jet launch as well as the jet and core ejecta properties, but is independent of the structure of the fast ejecta tail. The $\gamma$-ray emission depends on all the parameters including of course the tail. In order to consider configurations that can also account for the early optical emission we use similar jet properties and the same core structure as in  \citealt{Kasliwal17} and vary only the tail structure. We verify that indeed the obtained optical/IR emission (using the same calculation method as in \citealt{Kasliwal17}) is consistent with the observed one. Below, for completeness, we describe the full configuration we simulated. 
 
Initially  at $t=0$ (defined as the merger time) we have a cold ejecta that expands radially. It is present from the base of the grid at $ r_{esc} = 4\times 10^8 \cm$ up to $ r_{max} = 5.2 \times 10^9 \cm $. The ejecta has an angular profile, where most of the mass ($ 75\% $) is near the equator at $ \theta > 1.0 \,\rad $, where $\theta$ is the angle with respect to the axis. The ejecta is divided also in the radial direction into two regions (with the same angular profile) - the main massive slow part that extends at $t=0$ up to $r_c=1.3 \times 10^9$ cm and a low-mass fast tail that extends at $t=0$ between $r_c$ and $5.2 \times 10^9$ cm. The density profile of the dense part is:
\begin{equation}
\rho_{\rm{c}}(r,\theta) = \rho_0r^{-2}(\frac{1}{4}+sin^3\theta)~,
\end{equation}
where $ \rho_0 $ is the normalization which is chosen for a total ejecta mass $ M_c = 0.1\msun $. The velocity profile of the core is
\begin{equation}
v_{\rm{c}}(r) = v_{c,max}\frac{r}{r_c}~,
\end{equation}
{where $ v_{c,max} = 0.2c $ is the maximal velocity of the core. The fast tail density profile has  a very steep power-law in $v$ between $v_{c,max}$ and $v_{ej,max}$ and its normalisation is chosen so its total mass is $M_e$. Where needed we add an exponential (in density)  transition layer between the core and the tail in order to have a continuous density profile.}
The jet is injected into the ejecta with a delay of $ 0.8\s $ for a total working time of $ 1\s $ and a total luminosity of $L_j=2.6 \times 10^{51} ~\rm{erg~ s^{-1}}$. The jet is injected with a specific enthalpy of 20 at an opening angle of $ 0.7\rad $ from a nozzle at the base of the grid with a size of $ 10^8\cm $.

  
We improve the resolution of the simulation in \cite{Kasliwal17} as follows. In the $ r $-axis we use 3 patches, the innermost one in the $r$-axis  resolves the jet's nozzle with 20 uniform cells from $ r = 0 $ to $ r = 2\times 10^8\cm $. The next patch stretches logarithmically from $ r = 2\times 10^8\cm $ to $ r = 2\times 10^{10}\cm $ with 800 cells, and the last patch has 1200 uniform cells to $ r = 1.2\times 10^{12}\cm $. In the $ z $-axis we employ two uniform patches, one from $ z_{beg} = 4.5\times 10^8\cm $ to $ z = 2\times 10^{10}\cm $ with 800 cells, and the second to $ z = 1.2\times 10^{12}\cm $ with 1200 cells. In total the grid contains $ 2020 \times 2000 $ cells, and the simulation lasts 40 seconds.


\subsection{Hydrodynamics}
\label{sec:hydro}

\begin{figure}
\centering
\includegraphics[width=.48\textwidth]{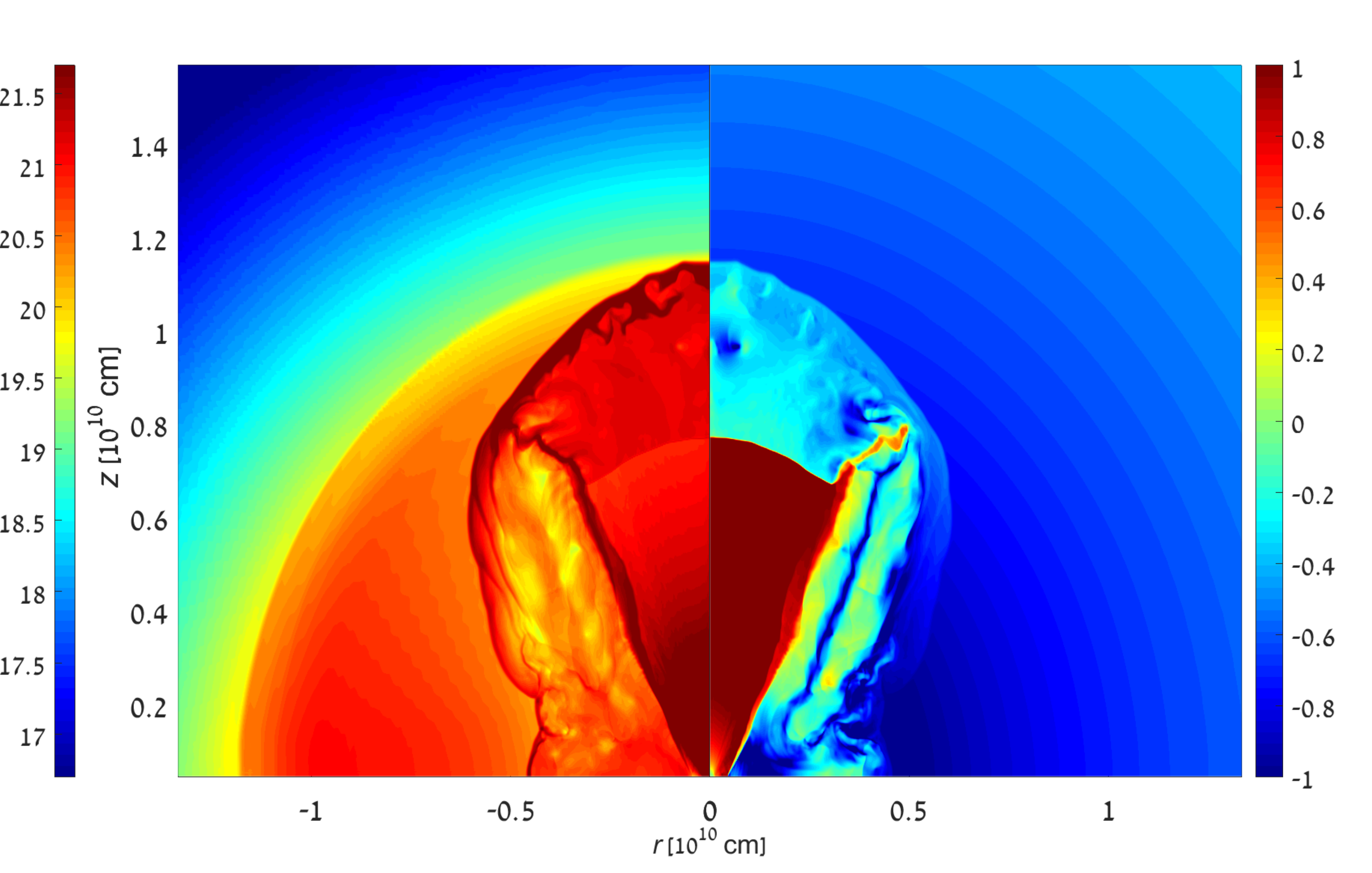}
\includegraphics[width=.48\textwidth]{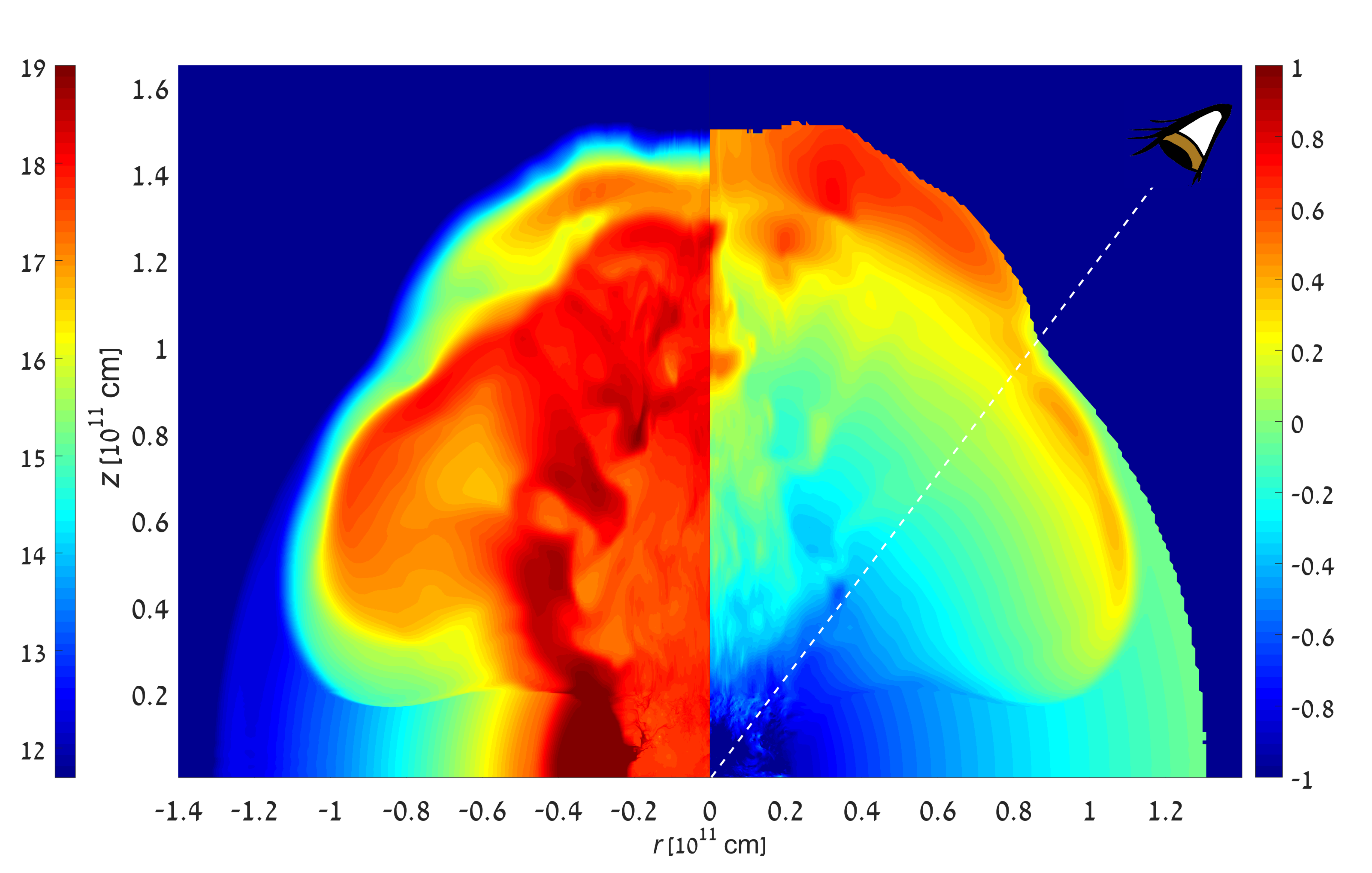}
\caption{Maps of the logarithmic energy density excluding the rest-mass energy (left) in c.g.s units and logarithmic four velocity (right). {The upper figure is taken before the breakout of the forward shock from the core ejecta. Although the forward shock will break out, the jet material behind the reverse shock will remain trapped inside and will be choked with the termination of the engine}. The lower figure is taken when the shock breaks out of the tail at $ \theta = 0.7 $rad at $ t = 6.2\s $ and $ r = 1.3\times 10^{11}\cm $. The shock has a quasi-spherical shape, reaching most of the ejecta. (An animation is available in the online journal.)} \label{fig:choked_map}
\end{figure}
 
At $ t = 0.8\s $ a jet is launched into the expanding ejecta, the jet is wide and covering a solid angle of about 25\% of the entire sphere. A large fraction of the shocked material accumulates on top of the jet head and cannot be evacuated as it is not in a causal contact with the jet outer envelope (see top panel in figure \ref{fig:choked_map}). The wide jet is not collimated, propagating roughly conically inside the core as it shocks a significant fraction of it.  After a total working time of $ 1\s $ the engine is turned off and within $ 0.5\s $ the jet is choked just before it emerges from the core ejecta depositing all the jet's energy into the cocoon. The cocoon then breaks out of the core into the low-mass tail. No emission is released yet to the observer because to the high optical depth of the tail, but due to its low density the cocoon expands sideways and accelerates into the tail, in a way that is almost similar to expansion in a vacuum.  First light is emitted upon the breakout of the cocoon from the fast ejecta tail (see bottom panel in figure \ref{fig:choked_map}). In the specific simulation depicted in figure \ref{fig:choked_map} the shock breakout at $ \theta = 0.7 $ takes place at $t=6.2\s $ at a radius of $ 1.3\times 10^{11}$ cm , corresponds to an observer time of $ \sim 1.8\s $ after the merger. At this point the shock is quasi-spherical and normal to the surface, crossing most angles at similar times, leaving only a fraction of unshocked ejecta around the equator. The velocity of the gas right behind the shock upon breakout is $ \Gamma \approx 2.0 $, but soon after the breakout it accelerates to $ \Gamma \approx 3.5 $.

\subsection{$\gamma$-rays}
\label{sec:gamma}

Turning now to our main results we consider the $\gamma$-ray emission of the cocoon's shock breakout. As mentioned earlier this emission depends on all the parameters including those of the  faster tail that surround the main ejecta. We kept the jet and core parameters constant and checked the effect of the tail by considering several configurations (without doing an exhaustive parameter phase space search). We examined tail parameters in the following ranges: the density power-law $-(5-15)$, total mass $ (10^{-4} - 5\times 10^{-2})\msun $ and maximal velocity $ (0.5-0.85)c$. 

{ The outcome depends only on the parameters near the shock upon breakout, which are determined by these initial conditions. The light curves we obtained showed a large range of observed values, yet almost all light curves showed the expected common features of low-luminosity (compared to the total ejecta energy), low variability and hard to soft evolution. For the range of parameters we considered we find a large variation in the luminosity, where the peak luminosity varies between $ 10^{46} ~\rm{erg~ s^{-1}} $ and $ 10^{49}~ \rm{erg~ s^{-1}} $. Most simulations have shown hard to soft evolution with two spectral components. The ratio between the peaks of the two component is typically a few and varies between simulations by about an order of magnitude. The peak energy of the hard component is typically a few hundred keV, but in extreme cases it exceeds 1MeV. The soft component is typically lower than 100 keV but it may go under 1 keV in extreme cases. Smaller variations are seen in the duration and the delay, where the observed duration varies between 0.5 s and 4 s and the delay with respect to the merger between 1.5 and 4 s. The shape of the light curve also varies. Most have a fast rise and slow decay, but some have the opposite behavior and some are symmetrical. Some curves showed more structure than others but none have shown a rapid variability. }

We find strong dependencies between the tail profile and the produced signal. First, a steeper tail density profile  leads to a stronger shock, which in turn produces a brighter {and harder} signal. Another important parameter is the tail's mass as the more massive ones stall the shock, filters less energetic material in high Lorentz factors, and result in a dimmer, later and longer signal. The tail's front velocity has several effects. One is the time and duration of the peak as the shock would experience an earlier breakout with slower ejecta velocities. However, in these cases, where the shock breakout takes place in relatively small radii, the shock is more oblique and hence weaker.
Additionally the shape of the signal varies greatly between the simulations, a lower ejecta front velocity mostly produces a slower rise and a sharp decline, while the fastest ejecta encounters a more spherical shock so that the angular time is shorter, the rise is sharp and the decay is gradual.
Finally, in most, {but not all}, simulations we have carried out, the spherical phase dominates over the planar, both in time and luminosity, giving rise to a stronger soft component in the spectrum.

\begin{figure}
\includegraphics[width=.5\textwidth]{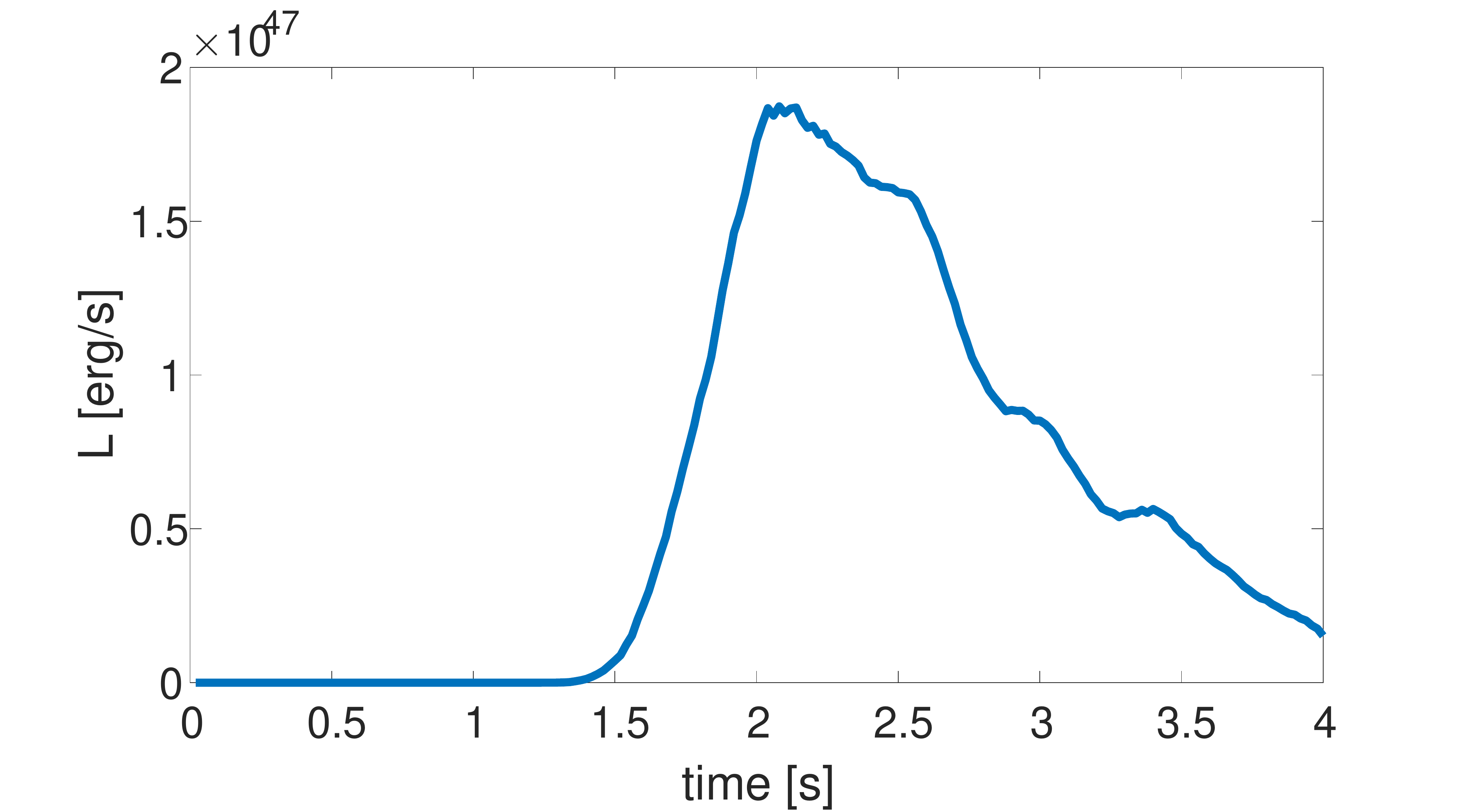}\\
\includegraphics[width=.5\textwidth]{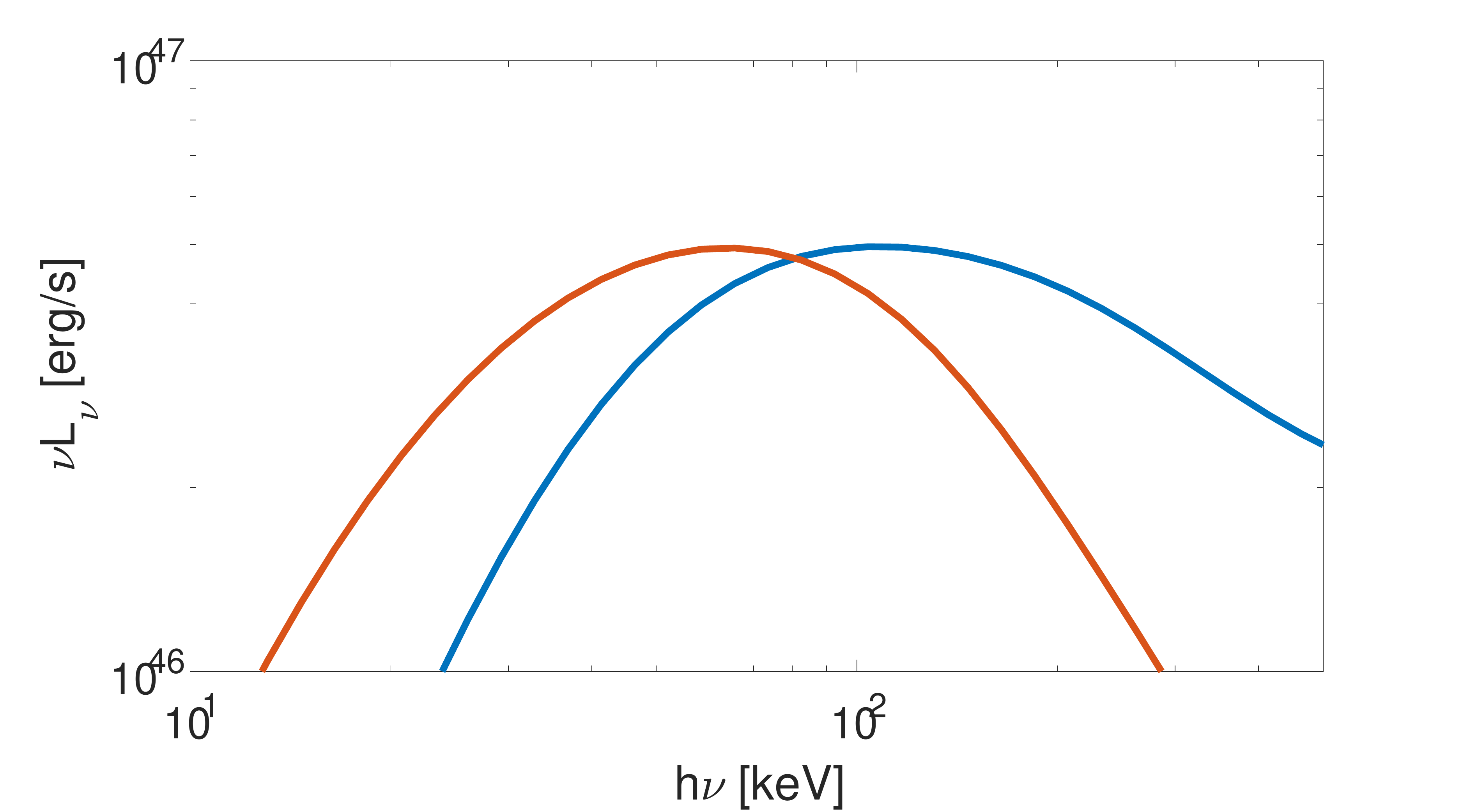}
	\caption { The light curve (top) and spectrum (bottom) of the $\gamma$-ray signal for an observer at $ \theta = 0.7 $. {The spectrum is divided at $ t = 2.5\s $ (earlier in blue and later in red), where the sharp drop in the luminosity begins.} In this configuration the tail density profile is a steep power-law of -14, and its velocity varies between $v_{c,max}=0.2c$, and  $v_{ej,max} = 0.7c $. The total mass of the tail is  $ 5 \times 10^{-3}\msun $, about 0.05 of the total mass of the ejecta. However,  only a small fraction of this mass,  $10^{-7}~  (10^{-8}) M_\odot$   has $v \ge  0.5~ (0.6)c$. }
	 \label{fig:lightcurve}
\end{figure}

Among the configurations that we have examined there were many light curves that showed characteristics, such as duration, luminosity and hard to soft spectral evolution, to within an order of magnitude compared to those observed in GRB 170817A. In figure \ref{fig:lightcurve} we show an example of the light curve and spectrum { observed  at $ 0.7\rad $} that agree exceptionally well with all the observed properties of GRB 170817A. 
The resulting light curve starts rising at 1.5 s after the merger. It peaks at 2 s after the merger at $1.8 \times 10^{47}$ erg s$^{-1}$, and its total duration is $T_{90}=1.75$ s. It is composed of a bright initial pulse that lasts less than a second. {It rises rapidly as the quasi-normal shock breaks out at $ \theta \approx 0.6 $. After the peak it drops  gradually due to the combination of the spherical phase and shock breakout at $ \theta \approx 0.8 $, before a short sharp drop following the end of the spherical phase at these angles, after which}  a more gradual component, originated from wider angles,  is observed for a bit longer than a second. The spectrum of the initial pulse is harder with $\nu L_\nu$ peaking at $110$ keV, where the tail peaks at $60$ keV, which corresponds to $T=15$ keV. We stress that the agreement that we find does not imply that the setup which has been used in this simulation is the one we expect that took place in GRB 170817A. Since we did not scan the entire parameter space systematically, we expect that there are many other setups which are able to produce a similar or even better agreement with the observations.

\subsection{UV/Optical/IR and Radio}
\label{sec:optrad}
\begin{figure}
			\includegraphics[width=.5\textwidth]{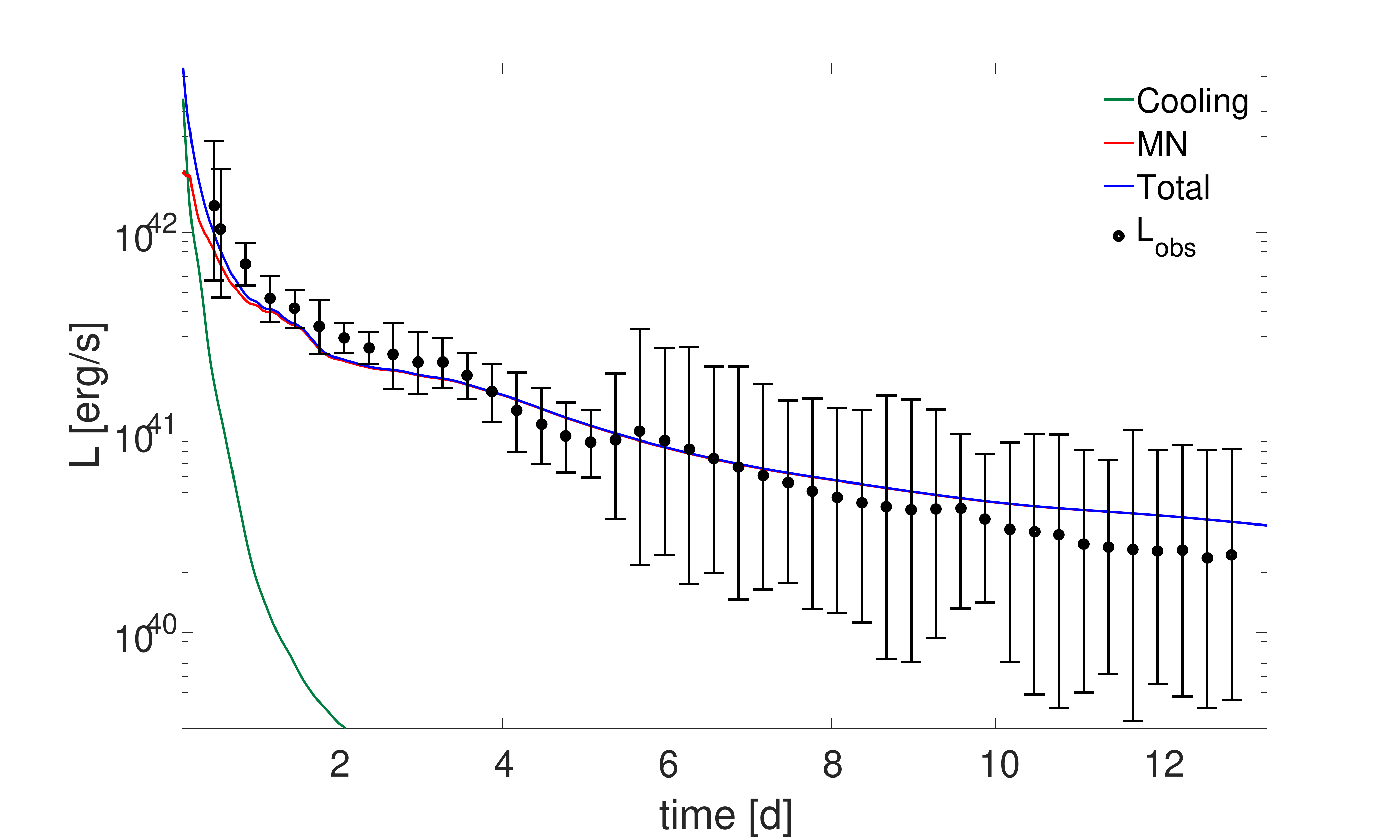}
			\includegraphics[width=.5\textwidth]{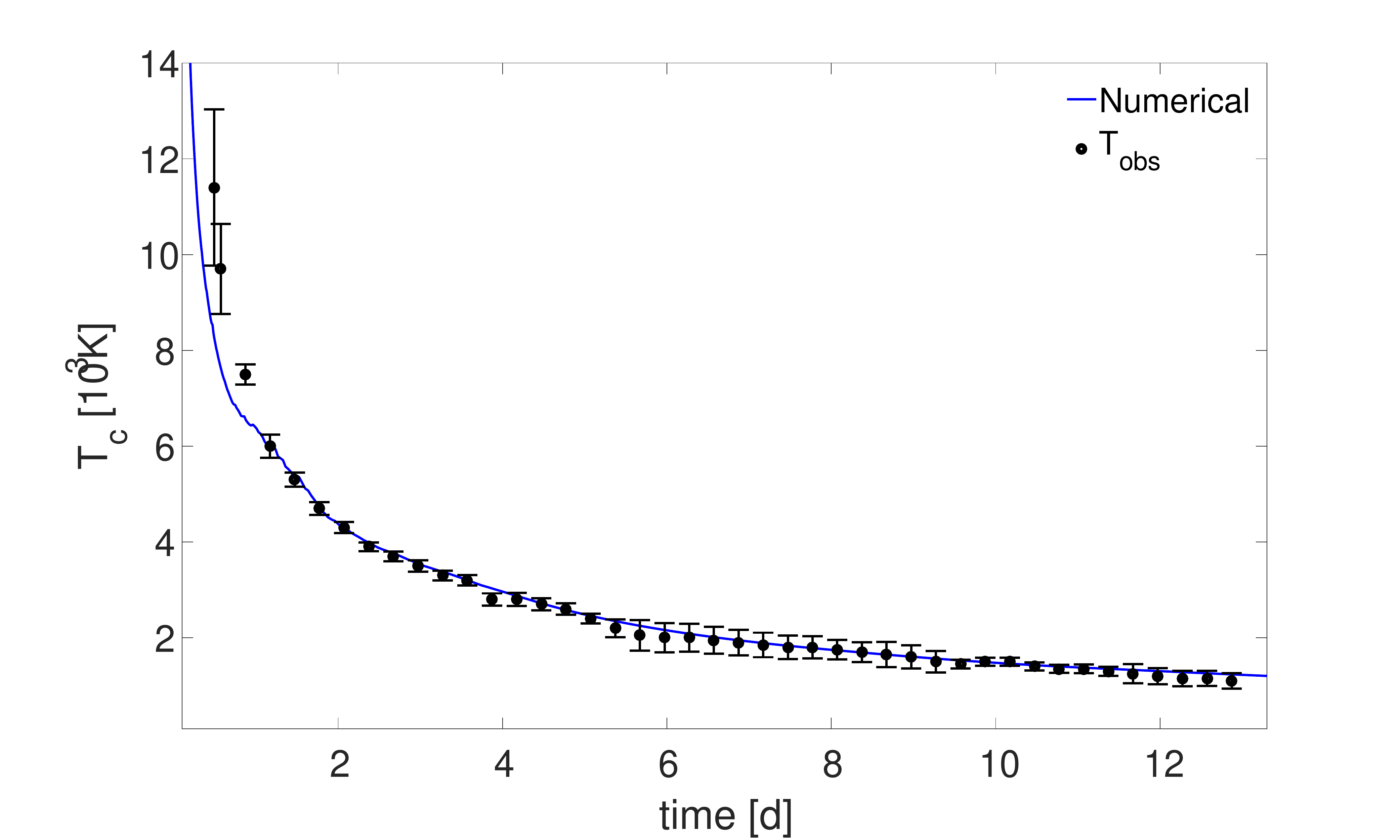}
	\caption { The UV/optical/IR macronova corresponding to the matter ejected in the choked jet case. Top: the bolometric UV/optical/IR light curve. Shown are the total emission (blue); cooling emission  (green); the cocoon macronova (red). Bottom: the temperature.  Observed data is from \citet{Kasliwal17}.  The simulation setup is the same as in figure \ref{fig:lightcurve}. {It is different from the setup of the simulation presented in \citet{Kasliwal17} only in the properties of the fast tail, and therefore the optical light curve here is very similar to the one presented there.} }\label{fig:opt}
\end{figure}

As discussed in \cite{Kasliwal17} the expanding ejecta produces a macronova signal powered by its radioactive decay, while its later interaction with the surrounding matter produces a radio afterglow \citep{Hallinan17}.  The macronova signal is boosted during the first day due to the mildly relativistic motion of the cocoon matter. The resulting optical/IR bolometric light curve and temperature, calculated from the same setup that produces the $\gamma$-rays depicted on figure \ref{fig:lightcurve} using the same methods described in \cite{Kasliwal17}, are shown in Fig. \ref{fig:opt}. Note that since we kept the same properties for the core of the ejecta as in \cite{Kasliwal17} and similar jet properties the optical emission from all our choked jet simulations results in Opt/IR light curve that is in general agreement with the observations. 

\begin{figure}
\includegraphics[width=.5\textwidth]{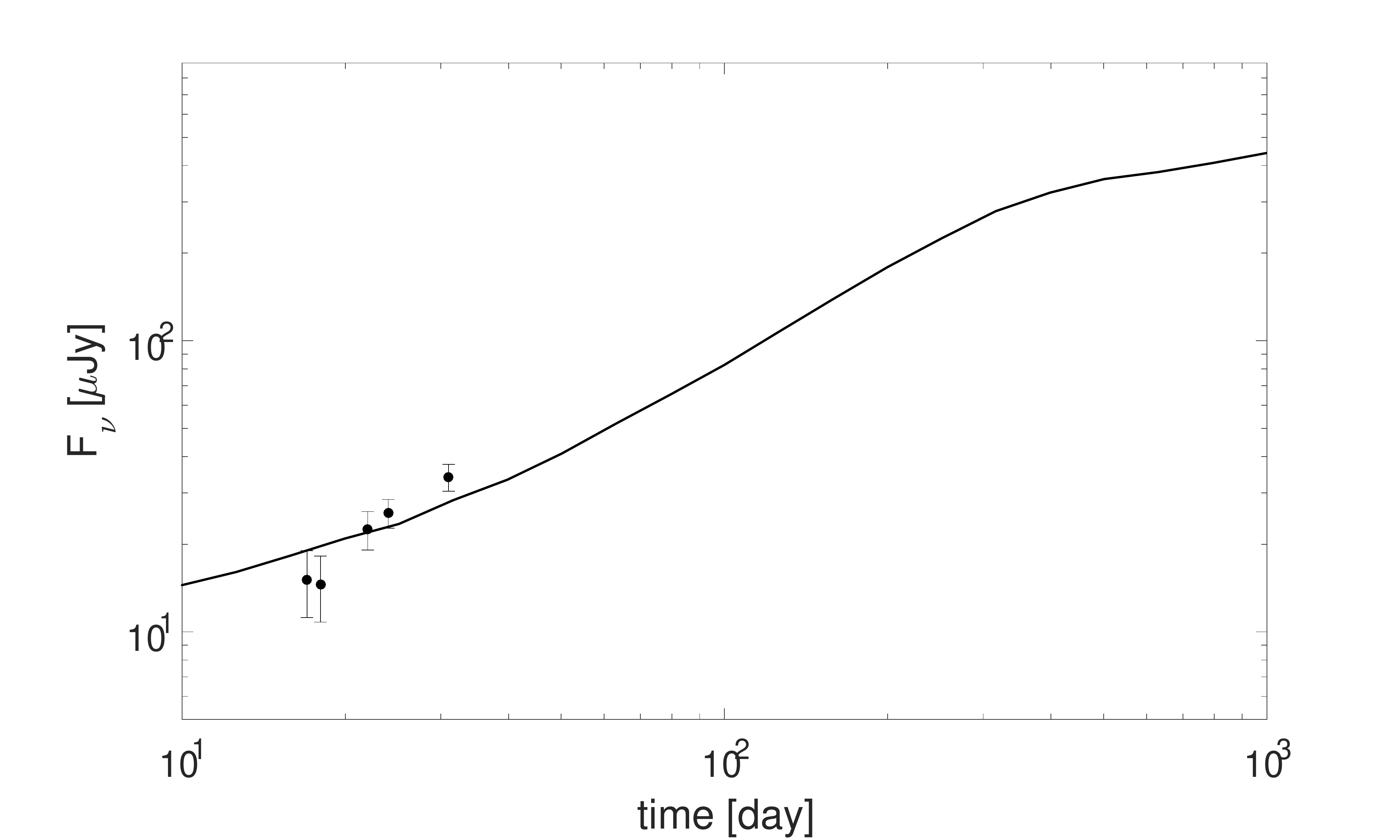}
	\caption { The radio light curve at 3 GHz corresponding to the matter ejected in the choked jet simulation used in figure \ref{fig:lightcurve}. The external density is uniform with $n=1.3 \times 10^{-4} ~{\rm cm}^{-3}$. The microphysical parameters we used are electron energy fraction $\epsilon_e=0.1$, magnetic field energy fraction $\epsilon_B=0.01$ and electron distribution power-law index $p=2.1$. 
	The observed points are from \citet{Hallinan17}. }
	\label{fig:radio}
\end{figure}

The interaction of the significant mildly relativistic outflow with the circum-merger material produces a strong radio afterglow signal \citep{nakar2011,piran2013,hotokezaka2015,Hallinan17}. This signal rises early because of the relativistic motion of the cocoon's boosted material. Fig. \ref{fig:radio} depicts the radio signal at 3 GHz, calculated from the same setup that produces the $\gamma$-ray signal depicted on figure \ref{fig:lightcurve}, and compares it with the observations form \citet{Hallinan17}. The radio light curve is calculated using the same method as in \citet{Hallinan17}. {The energy distribution as function of velocity  of the outflow (i.e., $E(>v)$) is  approximated to be spherically symmetric and is taken from the final snapshot of the simulation in the following way. For each Lorentz factor $\Gamma$ we measure all the energy that is within an opening angle of $1/\Gamma$ with respect to the line of sight and take its isotropic equivalent. } 
Note that we used the profile of the expanding material obtained from a simulation that was chosen based on its $\gamma$-ray signal without any fit to the radio. The only free parameter we used in fitting the radio is the circum-merger density, In general the radio signal is very sensitive to the exact velocity profile of the ejecta and each simulation results in a different prediction for the radio. Some are in better agreement with the observations and some in worse. An interesting prediction of the choked jet model, which are seen in all our simulations, is that the signal continuous to rise in the near future. 

\section{Successful Jet} 
\label{sec:succseful}
We turn now to our simulation of a successful jet. 
\cite{g17} have shown that the cocoon of a successful jet also generates a mildly relativistic outflow at a wide angle \footnote{Recently \cite{lazzati2017b} also found a similar component, however in their model they do not identify an energy dissipation mechanism that will produce the observed $\gamma$-rays.}. They show that following the breakout the cocoon accelerates for a short time and after doubling its radius it sets into a homologous expansion. The cocoon cools adiabatically loosing most of its internal energy and if there is no external source to dissipate the energy of the cocoon the remaining energy is  released once the optical depth drops at frequencies far below $\gamma$-rays. We examine here a configuration that is similar to the one explored in \cite{g17} to which a fast low-mass tail has been added. We show that  such a fast low-mass tail has a minor influence on the propagation of the cocoon once it emerges from the core of the ejecta. However, like in the case of a choked jet the breakout of the shock formed by the cocoon from the low-mass fast tail  produces a $\gamma$-ray signal comparable to the observed $\gamma$-rays from GRB 170817A. 

\cite{g17} have shown that  jet  propagation in the core of the ejecta must be done in 3D. We have used therefore  simulation $ \B $ of \cite{g17} up to the time the jet breaks out of the core. {Since following the evolution to the radii that are needed in order to explore the shock breakout from the fast tail is too demanding for our 3D simulation we map the system at the time the jet breaks out of the core from 3D to 2D and continue simulating the jet propagation within the extended fast tail in 2D. We verify that this does not strongly affect the results (see below)} Unlike the choked jet case we did not attempt to find a configuration that produces $\gamma$-rays with characteristics that are similar to those observed in GRB 170817A. Instead, we preformed only a single simulation which shows that also a cocoon breakout from a successful jet generates a $\sim 1$ s $\gamma$-ray signal with the correct delay which is bright enough and shows the hard to soft spectral evolution. We note, however, that this scenario has difficulties explaining the early  UV/Optical/IR signal as the cocoon's macronova in this case decays after several hours {(see also \citealt{g17})} and it is not powerful enough to account for the bright UV/blue signal observed one day after the burst. 
 
\subsection{Initial conditions and numerical setup}
 
The simulation is composed of two parts, the first, which follows the system up to the point the jet breaks out of the core ejecta is simulation $ \B $ of \cite{g17}, which is in 3D. The second is its later evolution in 2D. 
For completeness we provide here the initial configuration of simulation $ \B $ of \cite{g17}.\footnote{ We use the freedom to scale numerical simulations (see \citealt{granot2012}) to scale the mass and length by five and three times the values used in \cite{g17}, respectively. }.
At $ t = 0 $, the time of the merger, the merger is surrounded by a cold ejecta {starting from $ r_{esc} = 1.3\times 10^8\cm $}. The total mass of the ejecta is $ 0.05\msun $, with $\rho \propto r^{ -3.5}$. The ejecta outer surface is located at $ r = 3.9\times 10^8\cm $ with a velocity of 0.2c. At $ t = 0.72\s $ a narrow jet with an opening angle of $ 10^\circ $ is injected into the system. The jet which has a specific enthalpy of 20 and total luminosity $ 6.7 \times 10^{50}\rm{erg~s^{-1}} $, reaches the ejecta surface at $ 9 \times 10^9\cm $ after another $ 0.72\s $.
From this time onwards, the jet evolution is insensitive to the system's dimensionality, as the jet has evacuated all the bulk mass in front of it so that the 2D numerical artifact of the ``plug"  \citep[see discussion in][]{g17} will not be present. We verify this in $ \S $ \ref{app:convergence}
We therefore utilize the snapshot of the 3D simulation at the time the jet breaks out of the core as our initial conditions for the 2D simulation. We convert the 3D results into 2D by averaging over rings along the rotation axis. Additionally we add to this snapshot the light, $ 2\times 10^{-3}\msun $, tail ahead of the core. The tail's  density profile is a power-law with $\rho \propto r^{-10}$ and its front velocity is $ 0.8c $, keeping the homologous profile by extending up to 4 breakout radii.
 
The numerical setup  (solver, equation of state etc.) is identical to the choked jet simulation. {The grids are however somewhat different as to reflect the earlier 3D simulations. }
The grid is divided into three patches in each axis, while the first two are identical to the original 3D simulation. 
The innermost patches are distributed uniformly in $ r $ (50 cells) and $ z $ (400 cells) axes, extending to $ 3\times 10^8\cm $ and $ 6\times 10^9\cm $, respectively. The $ z $-axis begins at $ 1.3\times 10^8\cm $. The second patches are logarithmic with 240 and 600 cells up to $ 9\times 10^{10}\cm $ and $ 1.2\times 10^{11} $ in $ r $ and $ z $ axes, respectively. The extension of the grid to include the ejecta tail is to $ 1.2\times 10^{12}\cm $ and $ 1.5\times 10^{12}\cm $ with 1200 and 1500 uniform cells in $ r $ and $ z $ axes, respectively. In total the simulation contains $ 1490 \times 2500 $ cells and lasts $ 50\s $.
 
\subsection{Hydrodynamics}
 
\begin{figure}
             \centering
             \includegraphics[width=.5\textwidth]{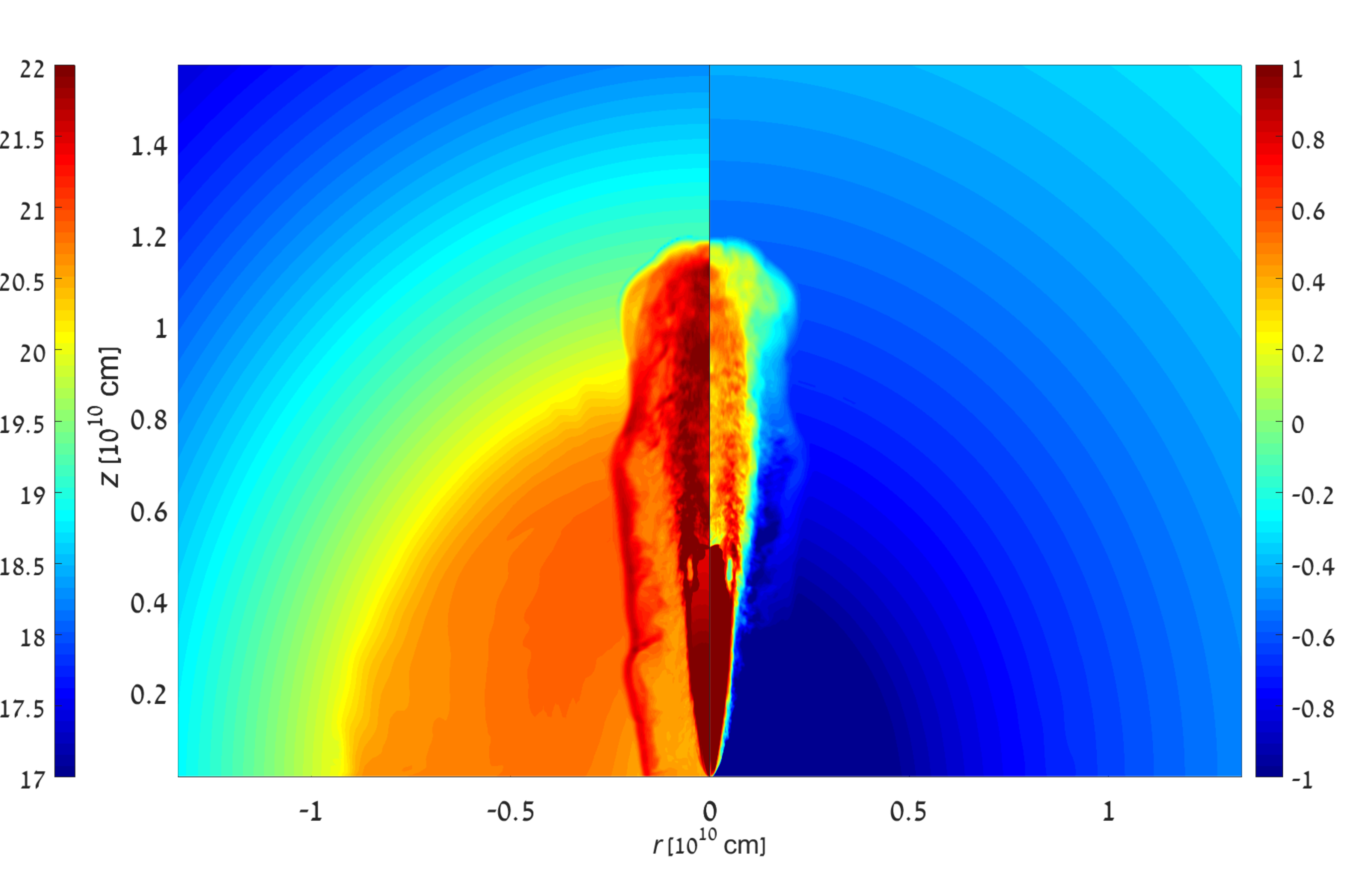}
             \includegraphics[width=.5\textwidth]{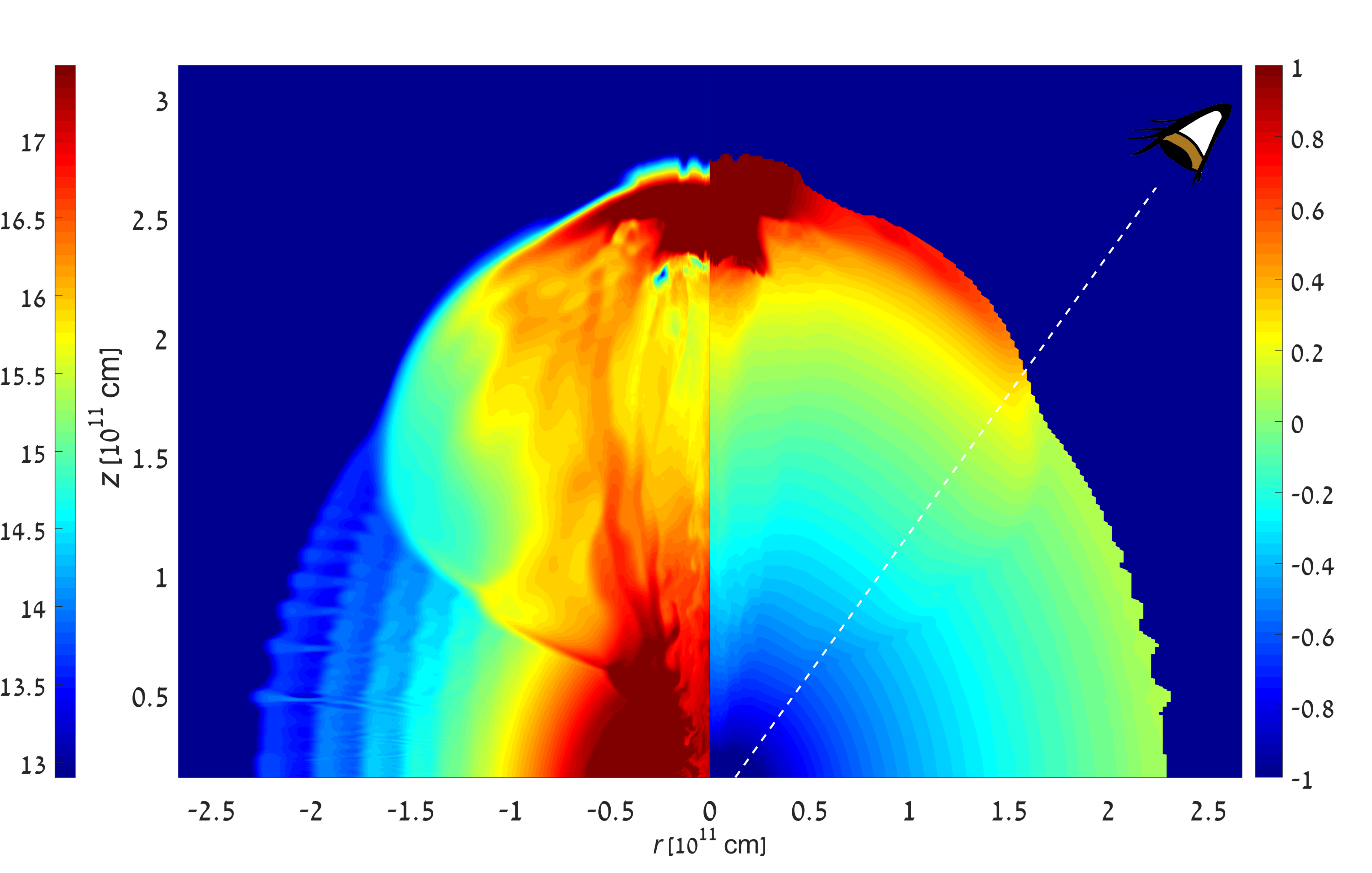}
             \caption{
            Same as Fig. \ref{fig:choked_map} for the successful jet case. Top: A well collimated jet reaching the edge of the core ejecta. Bottom:  The shock breakout at $ 0.7\rad $ after $ 9.8\s $ and at $ r = 2.4\times 10^{11}\cm $. The core ejecta near the origin and the jet at the top are prominent as the most energetic parts. The expanding structure of the cocoon with an oblique shock is clearly seen . (An animation is available in the online journal.)
               }
               \label{fig:jet_map}
\end{figure}

We have injected at $ 0.72\s $ after the merger, a narrow ($ \theta_j = 10^\circ $) jet  into the expanding ejecta {(Fig. \ref{fig:jet_map})}. The jet is well collimated and able to evacuate efficiently the ejecta in front of it and propagate at mildly relativistic velocities until breaking out of the core ejecta within another $ 0.72\s $, before its engine is turned off $ 1\s $ after the launch.
At this point the jet enters the dilute extended tail, and accelerates to a Lorentz factor of a couple of dozens. The jet is accompanied by a hot cocoon that expands to a wide angle and moves in mildly relativistic velocities. The cocoon shape is aspherical, and the shock breakout is oblique. It does not reach angles larger than $ \pi/4 $. However it is fast, and its Lorentz factor is almost 3 upon breakout and 5 after the acceleration phase.
 
In Fig. \ref{fig:jet_map} we show the breakout at $ \theta = 0.7\rad $. It takes place after $ 9.8\s $ and at $ r = 2.4\times 10^{11}\cm $, corresponds to $ t_{obs} \approx 1.8\s $. The main differences from the choked jet case (Fig. \ref{fig:choked_map}) are {the initial jet collimation, (shown in the top panel) and its presence in the homologous phase on the $ z $-axis (shown in the bottom panel) at the time of the cocoon breakout} with a width of slightly more than a light second, and the cocoon which is less spherical.

\subsection{$ \gamma$-rays }
 
\begin{figure}
\centering{
{\includegraphics[width=.5\textwidth]{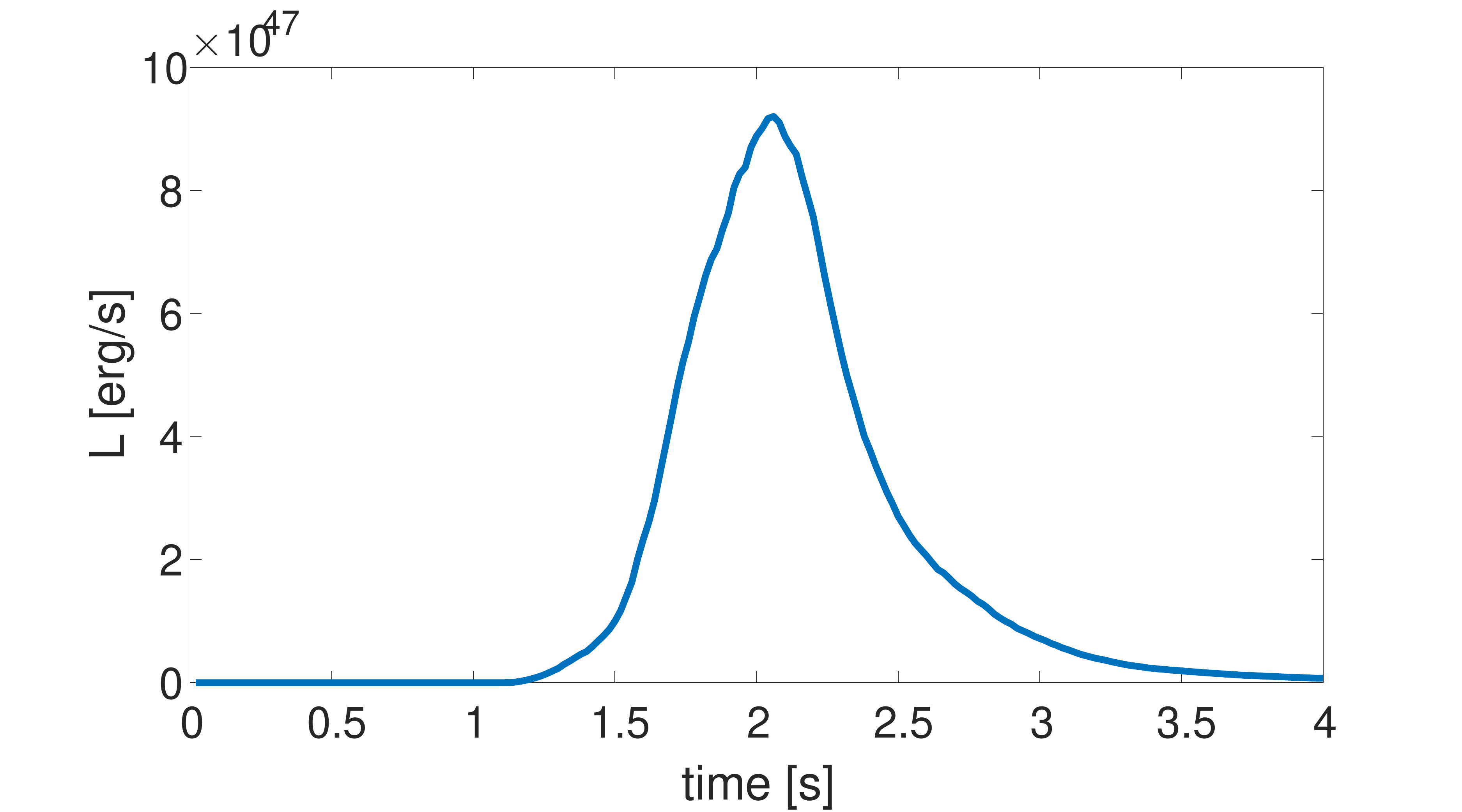}}\\
{\includegraphics[width=.5\textwidth]{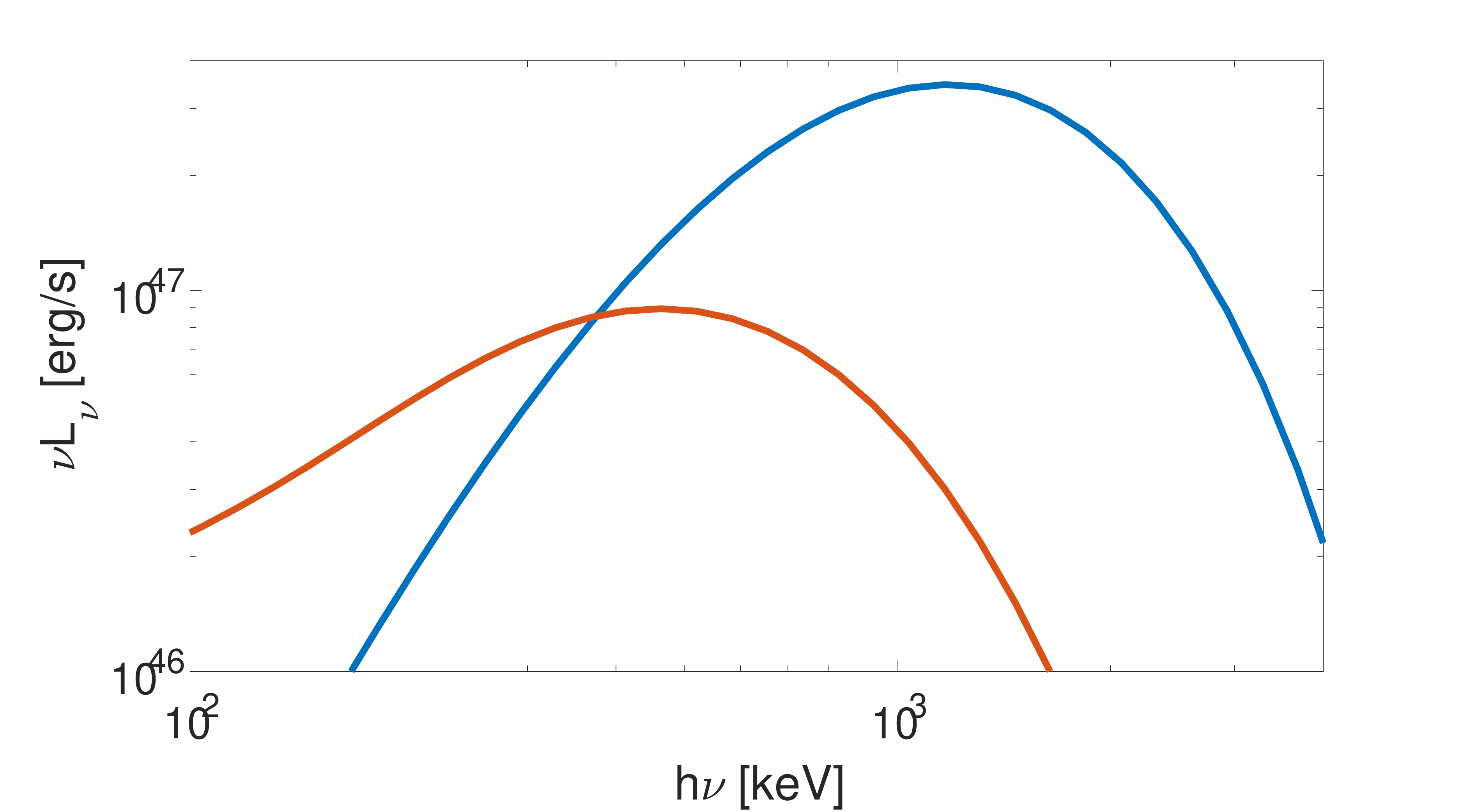}}
\caption {The $\gamma$-ray light curve (top) and spectra (bottom) of the successful jet simulation that we carried out as seen by an observer at $ \theta_{obs} = 0.7\rad $. The spectrum, which is divided at $ t = 2.3\s$, shows two component, early hard (blue) and late soft (red). }
\label{fig:jet_lightcurve}}
\end{figure}

{We calculate the $ \gamma $-ray emission arising from a shock breakout of the cocoon from the extended tail at large angles, where the emission from the jet itself does not contribute at all. In Figure \ref{fig:jet_lightcurve} we present the signal for an observer at $ \theta_{obs} = 0.7\rad $.  The delay of slightly less than two seconds, and the duration $ T_{90} = 1.6\s $ are similar to GRB 170817A. The light curve shape in this simulation is determined mostly by the obliqueness of the shock. The fast rise to the peak is due to the shock at $ 0.55 < \theta < 0.7 $, the peak is maintained by angular contribution from the shock at $ 0.4 < \theta < 0.55 $, followed by a steep decline as the shock does not reach angles larger than $ \pi/4 $. 
With a peak luminosity of   $ 9\times 10^{47} \rm{erg~s^{-1}}$ the signal from the simulation is brighter by about an order of magnitude compared to GRB 170817A.  The spectrum shows a clear hard to soft evolution, but both components are harder by an order of magnitude compare to GRB 170817A.  Dividing the spectra at $ t = 2.3\s $, during the sharp drop from the peak of the signal, the hard component is about 1MeV, while the softer one is several hundreds of keV.}
 
Given that (i) we have used an existing 3D  simulation as our initial condition and (ii)  we did not do any parameter search but we run only a single set of parameters for the extended tail, the fact that most features of GRB170817 are present and fit up to better than an order of magnitude with this model is exceptional. Scanning the parameters space carefully is most likely to yield a significantly better match with the observed $ \gamma $-ray signal of GRB 170817A.
 
\section{Dissociation of the nuclei}
\label{sec:photodissociation} 

Before concluding we note a possibility, that we did not consider in this work, which is  the dissociation of heavy nuclei 
by the cocoon's  shock.  The internal energy per baryon in the shocked ejecta
is $\gsim 10$ MeV. This  exceeds the binding energy per baryon of heavy nuclides $\sim 8$ MeV.  In principle there 
is enough energy to 
dissociate  heavy nuclei in the shocked ejecta. As most of the energy is stored in the photons, 
the condition of the photodissociation 
of heavy nuclei depends sensitively on  the temperature.  The reaction rate is proportional
to the number of photons above $\sim 8$ MeV, which is strongly suppressed in the exponential tail of the photon's spectrum.
{At thermal equilibrium, which is expected at the shocked ejecta,} the critical temperature, above which  photodissociation occurs, is $\sim 200$ keV (e.g., \citealt{Woosley1978}).

{The jet propagates sub-relativistically within the core ejecta. Assuming that the jet is not strongly collimated (as in case of a wide jet) the velocity of the forward shock driven into the ejecta in its own frame is $\beta_s' \sim [L_{j,iso}/(4\pi R^2 \rho_{ej} c^5)]^{0.5} \sim 0.03 L_{j,iso,52}^{1/2}(M_ej/0.1\msun)^{-1/2} (R/10^9 {\rm cm})^{1/2}$ where $L_{j,iso}$ is the jet isotropic equivalent luminosity \citep{bromberg2011}. The  temperature behind the shock is estimated as $T \sim (\rho_{ej} v_s'^2/a_{BB})^{1/4} \sim 150 L_{j,iso,52}^{1/4} (R/10^9 {\rm cm})^{-1/2} keV $. If the jet is collimated than it propagates faster and the temperature is higher. This implies, giving the high sensitivity on the temperature, that photodissociation is marginal and depends mostly on the jet luminosity and the delay between the collapse and the launch of the jet. A powerful jet with a short delay ($\lesssim 0.1 s$) will dissociate nuclei in the ejecta along its path, while a less powerful jet and/or a longer delay will prevent photodissociation. In our simulations of choked and successful jets the delay is almost 1s so the jet interacts with the ejecta at radii $>10^9$cm 
and the typical temperature is $\approx 100$keV, thus we do not expect photodissociation to take place. We stress, however, that as this process is sensitive to the temperature and density evolution, the final abundance of nuclei should be addressed with more detailed calculations.}

In addition to photodissociation, free neutrons may also play an important role to disintegrate heavy nuclei. In the shocked material, free neutrons from the upstream have a velocity of $\mathcal{O}(0.1c)$. They thermalize through collision with heavy nuclei. During this thermalization process, heavy nuclei are disintegrated. This process occurs until a few hundreds ms after the mass ejection, by which free neutrons are likely exhausted for the typical ejecta parameters ($v_{ej}<0.5c$). Note that free neutrons can remain in the fast components of the ejecta with $v_{ej}\gsim 0.5c$ \citep{Metzger2015}.
Energetic free neutrons in the shock disintegrate heavy nuclei within the fast tail even at later times. 

{To conclude the jet propagation may affect the ejecta composition. This effect is only along the jet path, so a wide choked jet may affect a larger portion than a narrow collimated jet. In any case most of the ejecta, mostly the part which propagates at low latitudes will not  be affected. If the jet dissociates heavy nuclei in high latitude ejecta it will change its composition. The heavy nuclei will disappear and this will reduce the opacity and affect the radioactive heating rate within the region influenced by the cocoon. This will clearly affect the UV/optical/IR macronova signal especially at early times. Such process may be related to the blue light observed during the first day. The composition, however, will not strongly influence the $\gamma$-rays that arise from the shock breakout or the late radio emission which arises due to the interaction of the ejecta with the surrounding matter. }
\section{Conclusions}
\label{sec:conclusions} 

The short GRB 170817A was a not a regular sGRB. It was a {\it llsGRB}, namely a low luminosity one. Like {\it ll}GRBs (low-luminosity long GRBs) that are not produced by the same mechanism as regular long GRBs, {\it llsGRBs} are not produced by the same mechanism as regular {sGRBs}. In fact we suggest that, while the astrophysical scenarios are very different  (a Collapsar vs a merger) there is a similarity between the physical mechanism that produces the two types of low-luminosity GRBs. Both are produced by a shock breakout. In the former the shock breakout from the envelope of the star. In the latter the shock breakout is from the surrounding matter (ejecta) that is thrown out to space during the merger process. 

We have first shown, using the traditional compactness argument,  that the observed $\gamma$-rays implied that the event involved at least mildly relativistic outflow. At the same time this was not a regular sGRB viewed from the side \citep{Kasliwal17}. There is no simple way to obtain a spherical relativistic outflow in the configuration involved. To accelerate spherically the outer layers of the ejecta to $\gamma \gsim 2$ we will need to accelerate the whole bulk of the ejecta to this velocity and this is inconsistent with the observations. Thus, we conclude that the system must have involved a relativistic jet that carries the energy through the bulk of the ejecta to the outer layers and deposits it there, accelerating only a very small fraction of the ejecta. This jet might have penetrated successfully the ejecta producing a sGRB that was pointing away from us (and not observed by us), or it might have been completely choked within the outflow. In either case the jet would have produced a hot cocoon within the ejecta and we suggest that the observed $\gamma$-rays arose during the breakout of this cocoon from the fast outer tail of the ejecta. 

We have outlined here {an order of magnitude model} for a relativistic shock breakout from a moving ejecta, highlighting the differences between a shock that emerges from a static medium to one that emerges from an expanding one. We have shown that three generic properties of such a shock breakout are: (i)  The light curve is smooth. {It may show some structure but no fast variability;} (ii) Only a small fraction of the total available energy is emitted at this stage;  (iii) The emission involves two phases, a planar phase with a hard spectrum and a spherical phase that has a softer thermal spectrum. These feature resembles well the observed features of the observed {\it lls}GRB 170817A. It is important to note that the shock breakout scenario has four parameters that control its emission: The Lorentz factor of the shock, $\Gamma_s$,  the velocity of the ejecta, $\beta_{ej,max}$, the radius in which the shock breakout takes place, $R_{bo}$  and the causally connected mass that is within $R_{bo}/\Gamma_s$. {Moreover, there is little freedom in the amount of this mass allowed by reasonable ejecta models, and the observables (e.g. the total energy) are rather weakly dependent on it.} These should be compared with five observables, the total energy, the duration, the delay after the merger, the peak energy of the early hard component and the peak energy of the  soft component. 
Thus, the model is over-constrained and the basic agreement  of its predictions with the observations points out in its favor. Note that the model  requires a fast moving low-mass tail surrounding the main ejecta. However, as pointed out in \S \ref{sec:model} such fast moving material is expected \citep{Kyutoku2014,Hotokezaka2018} and was possibly noticed in some numerical simulations \citep{Bauswein2013, Hotokezaka2013}.

We then carried out numerical relativistic hydrodynamics  simulations that follow the propagation of a jet through the expanding ejecta, the formation of a cocoon and the breakout of the shock produced by the cocoon from the ejecta. We  post-processed the hydrodynamic simulation to calculate the resulting $\gamma$-ray emission. For the choked jet scenario we kept the parameters of the jet and the main component of the ejecta {(its massive core)} and have explored a small part of the phase space of fast tail configurations. Remarkably we have found over a large fraction of the  phase space that we explored  a $\gamma$-ray signal that is comparable {to within an order of magnitude} with the one observed in GRB 170817A. 

{The observed associated macronova \citep[e.g.][]{Kasliwal17} was also somewhat different from the expectations. It had a strong  early UV/Opt signal that was much brighter and bluer than earlier expectations. \cite{Kasliwal17} showed that a cocoon from a choked jet can also account for this early blue light. This is mostly due to a relativistic boost given at early times by the fast cocoon material. This boost makes the macronova signal both brighter and bluer, before the emission of the slower ejecta dominates at later times. All our {choked jet} simulations here, which have similar jet and ejecta core parameters as those of \cite{Kasliwal17},  recover the observed macronova. The mildly relativistic cocoon material interacts later with the surrounding matter and produces a radio signal. This signal is sensitive to the velocity distribution of the cocoon, which in turn depends on all the jet and ejecta parameters. We calculated the radio emission predicted for one of our simulations, that fit both the $\gamma$-rays and the macronova, and found that it is also in general agreement with the observed radio emission \citep{Hallinan17}. We find it remarkable that within a single set of parameters we find agreement with the observed $\gamma$-rays, macronova and radio signals. A clear prediction of the choked jet scenario is that the radio signal will keep increasing over the near future. }

While we have focused on the case of a choked jet, due to its ability to explain also the macronova, we have also shown that a rather similar $\gamma$-ray signal can be generated by a cocoon shock breakout driven by a jet that emerges and produces a sGRB pointing away from us. Here we have studied only one configuration whose $\gamma$-ray features resemble the observations. {A successful jet produces a much less massive and energetic cocoon and is therefore not expected to affect the macronova emission for more than several hours after the merger. Its lower energy also predicts a fainter radio signal which may be dominated by the off-axis radio afterglow of the relativistic jet.}

{We also examined briefly the effect of the jet propagation on the ejecta composition. We find that the temperature in the shocked core ejecta is marginal for photodissociation of the heavy nuclei. A luminous jet ($L_{j,iso} \gtrsim 10^{52}$ erg/s) that is launched quickly after the merger (within $\sim 0.1$ s) is expected to dissociate a significant fraction of the ejecta nuclei that lie in its path, while a less luminous jet and/or a longer delay will most likely have a minor effect on the ejecta composition. In the scenarios we considered here the jet was launched  about 1s after the merger and thus the temperatures found in our simulations are too low for photodissociation. It is possible, however, that for other events, or even for this one, the conditions are such that photodissociation takes place. If it does then it may strongly affect the early optical emission (which may be related to the observed first day blue light), but not the late optical/IR emission nor the $\gamma$-rays or the radio.}

Finally we note that while the cocoon breakout $\gamma$-ray signal  is much wider than the emission of a regular sGRB (in both cases of choked and successful jets), it is not isotropic. { The signal depends on the observer angle and nearer to the jet's axis it is typically brighter, harder, shorter and with a smaller delay} (clearly in the case of a successful jet the sGRB emission from the jet dominates completely over this emission for an on-axis observer). It is dimmer for larger observing angles (a detailed study of the emission as a function of the viewing angle will be published elsewhere), yet for some parameters it may be observed over the entire $4 \pi$. Overall we expect that the {\it lls}GRB signal will  be observed over a quite large solid angle and will accompany many GW events.

This research was supported by the I-Core center of excellence of the CHE-ISF. OG and EN were partially supported by an ERC starting grant (GRB/SN) and an ISF grant (1277/13). TP was partially supported by an advanced ERC grant TReX and by a grant from the Templeton foundation.  TP acknowledge kind hospitality at the Flatiron institute while some of this research was done. KH  is supported by the Lyman Spitzer Jr. Fellowship.


\appendix

\section{The $ \gamma $-ray shock breakout emission}\label{app:Calculation}
To calculate  the $ \gamma $-ray emission from the shock breakout we begin by monitoring the shock location as a function of time and angle. Then, for each angle $ \theta $ we find the shock breakout time $ t_{bo}(\theta) $ and radius $ r_{bo}(\theta) $.
For each lab frame time $ t > t_{bo}(\theta) $, the emission from angle $ \theta $ originates in $ r_e(\theta) $, the radius from where  photons diffuse out to the photosphere ($ r_{ph}(\theta) \equiv r(\theta,\tau=1)$) during the time since the shock crossing. We assume that the photons diffusion to be radial. Hence the location $ r_e(\theta) $ is determined by
\begin{equation}
r_e(\theta) = r_{ph}(\theta) - \frac{t-t_{bo}(\theta)}{\Gamma(r_e)^2}\frac{c}{\tau(r_e)}~,
\end{equation}
where $ \Gamma(r_e) $ is the Lorentz factor of the emitting region.
The released rest-frame energy per solid angle $ d\Omega $ between two successive time-steps $ t_1, t_2 $ is the total thermal energy $ 4p $ of the emitting region:
\begin{equation}
\frac{dE'(t_1,t_2,\theta)}{d\Omega} = \int_{r_{e}(t_1)}^{r_e(t_2)} 4p(r)\Gamma(r) r^2 dr,
\end{equation}
where $r_e(t_{bo}(\theta))=\infty$.
The observed energy is then obtained by boosting the rest-frame energy $ dE' $ to the observer, and the arrival time is given by considering the light-travel time from the photosphere. The observed bolometric luminosity is then calculated by integrating over all times and angles.

\section{Convergence test}\label{app:convergence}
\begin{figure}
	\centering
	\includegraphics[width=.48\textwidth]{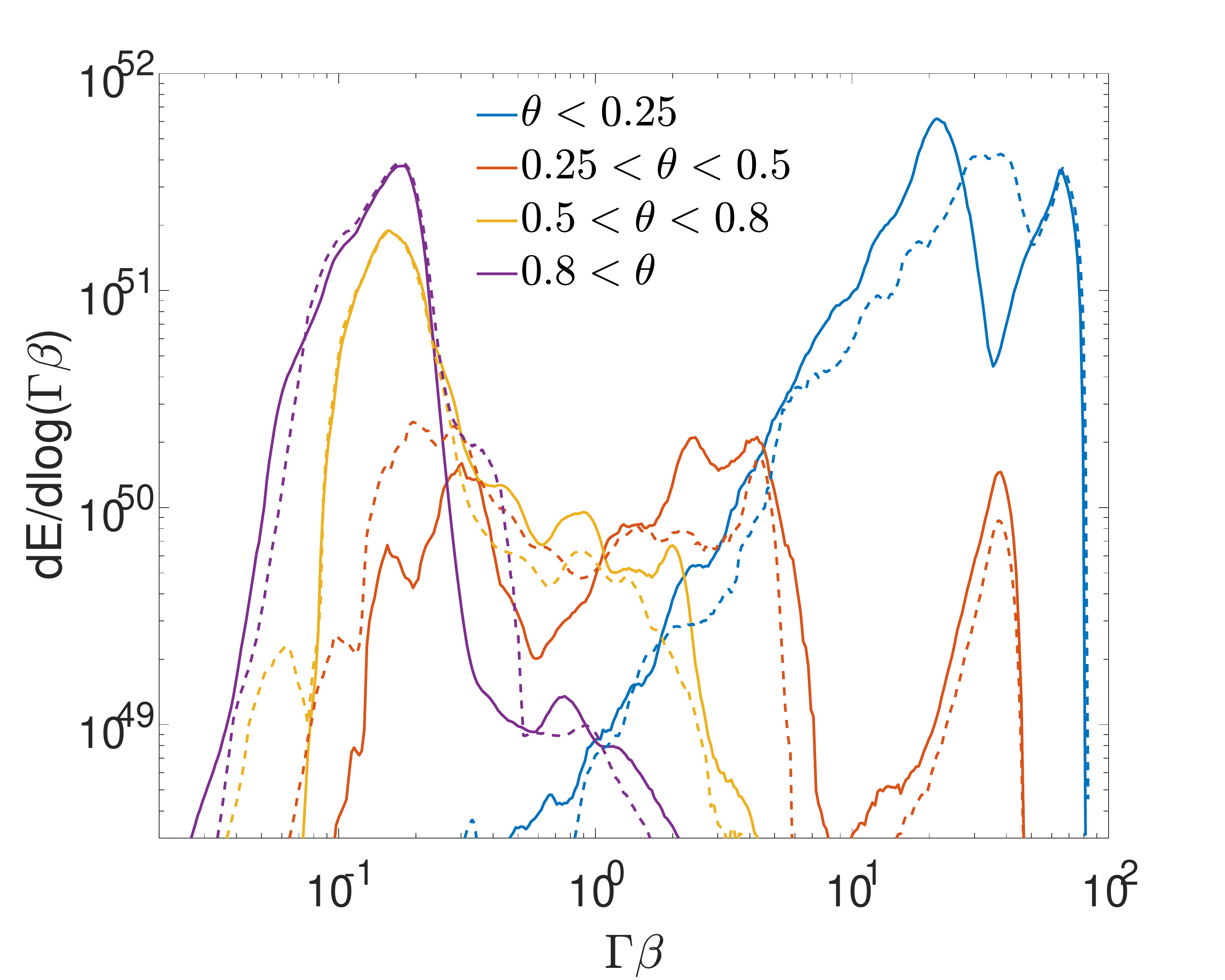}
	\caption{{Energy distributions per a logarithmic scale of $ \Gamma\beta $ for various ranges of angles in 3D (solid) and 2D simulations of the successful jet case, when no tail ejecta is present. The 2D simulation starts at the time that the jet breakout of the core ejecta and its initial conditions are a mapping of the 3D simulation at this point. The energy distributions in the figure are taken once the jet reaches a radius that is 10 times the breakout radius.
		}} \label{fig:convergence}
\end{figure}

Computer time limitations make it impossible to carry the full simulations in 3D. Therefore, in the collimated jet simulation, where 3D are necessary as long as the jet is collimated,  we switch from 3D to 2D when the configuration becomes insensitive to 3D effects, that is once a successful jet breaks out from the ejecta. We verify that converting the 3D simulation of the successful jet to 2D upon breakout from the core ejecta does not heavily affect the simulation outcome by performing two tests. 
First, we verified that a 2D simulation with a low-mass tail and one without one give similar results in their energy distributions at different angles.
Then, we run identical 2D and 3D simulations of the jet in a setup where there is no fast tail to the ejecta, starting at the time the jet and the cocoon break out of the core and until the jet increases its radius by a factor of 10 (initial conditions are taken from the 3D simulation). In figure \ref{fig:convergence} we compare the 3D and 2D velocity and energy distributions at various angles and found them to have a high degree of similarity at the last snapshots of the two simulations. The biggest discrepancy is fount in small velocities, around $ \Gamma\beta = 0.1 $, where the material has not reached the homologous phase yet. This component  does not contribute to the $ \gamma $-ray emission.

\label{lastpage}
\end{document}